\documentclass[12pt]{article}
\usepackage{graphics,graphicx,psfrag,epsf}
\usepackage{enumerate}
\usepackage{natbib}
\usepackage{amssymb, amsmath, latexsym, amsfonts, amsopn, epsfig, theorem, mathrsfs, booktabs, MnSymbol, multirow, color}

\newcommand{\blind}{1}

\def\R{\mathbb{R}}

\def\N{\mathbb{N}}
\def\P{\mathbb{P}}

\def\PP{\mathbb{P}}
\def\BE{\mathbb{E}}
\def\BV{\mathbb{V}}

\def\cd{\stackrel{\mathcal{D}}{\longrightarrow}}

\def\cas{\stackrel{\mbox{\small{a.s.}}}{\longrightarrow}}

\def\diag{\mbox{diag}}
\newcommand{\edist}{\stackrel{\mbox{\scriptsize${\cal D}$}}{=}}

\newtheorem{theorem}{\bf{Theorem}}[section]

\newtheorem{propos}[theorem]{Proposition}

\newtheorem{coro}[theorem]{Corollary}

\begin{document}

\def\spacingset#1{\renewcommand{\baselinestretch}%
{#1}\small\normalsize} \spacingset{1}


\if1\blind
{
  \title{\bf Goodness-of-fit tests for complete spatial randomness based on Minkowski functionals of binary images}
  \author{Bruno Ebner, Norbert Henze,
    Michael A. Klatt\thanks{The author gratefully acknowledges funding by the German Research Foundation (DFG) via the Grants No. HU1874/3-2 and No. LA965/6-2 awarded as part of the DFG-Forschergruppe FOR 1548 ``Geometry and Physics of Spatial Random Systems''.}\\
    Institute of Stochastics, Karlsruhe Institute of Technology (KIT),\\
    Englerstr. 2, D-76131 Karlsruhe, Germany\\
    and \\
    Klaus Mecke\\
    Institute of Theoretical Physics 1, University Erlangen-N\"{u}rnberg,\\
    Staudtstr. 7, D-91058 Erlangen, Germany}
  \maketitle
} \fi

\if0\blind
{
  \bigskip
  \bigskip
  \bigskip
  \begin{center}
    {\LARGE\bf Goodness-of-fit tests for complete spatial\\ randomness based on Minkowski functionals\\\bigskip of binary images}
  \end{center}
  \medskip
} \fi

\bigskip
\begin{abstract}
We propose a class of goodness-of-fit tests for complete spatial randomness (CSR). In contrast to standard tests, our procedure utilizes a transformation of the data to a binary image, which is then characterized by geometric functionals. Under a suitable limiting regime, we derive the asymptotic distribution of the test statistics under the null hypothesis and almost sure limits under certain alternatives. The new tests are computationally efficient, and simulations show that they are strong competitors to other tests of CSR. The tests are applied to a real data set in gamma-ray astronomy, and immediate extensions are presented to encourage further work.
\end{abstract}

\noindent%
\vfill

\newpage
\spacingset{1.45} 
\section{Introduction}
The statistical analysis of spatial point pattern data in a given study area $S$ (often called the observation window) is a classical task in many applications, including biostatistics
(e.g. structure analysis, \cite{25}), astronomy (e.g. detection of gamma-ray sources, \cite{24}), military (e.g. mine field detection, \cite{22}) and medicine (e.g. cluster detection in leukemia incidence, \cite{23}). A main objective is to characterize possible departures from so-called complete spatial randomness (CSR) of point patterns, which characterizes the absence of structure in data. CSR models the non-occurrence of dependence of point events within a given study area $S$, and it is synonymous with a homogeneous spatial Poisson point process (PPP). For an introduction to the concept of CSR we refer to \cite{15}, Section 8.4, and \cite{17}, Chapter 8. To be precise, we model the observed data by
\begin{equation*}
{\cal P}_\lambda := \{X_1,\ldots,X_{N_\lambda}\},
\end{equation*}
where $\lambda>0$, $(X_j)_{j\ge1}$ is a sequence of independent identically distributed random vectors taking values in $S$, and $N_\lambda$ is a nonnegative integer-valued random variable, independent of $(X_j)_{j \ge 1}$, with a distribution  that depends on some parameter $\lambda>0$. All random elements are defined on the same probability space $(\Omega,{\cal A},\P)$.
For ${\cal P}_\lambda$ to be CSR the $(X_j)_{j\ge1}$ are uniformly distributed on $S$, and $N_\lambda$ has a Poisson distribution with expectation $\lambda$.
The assumption of CSR will be called the {\em null hypothesis} $H_0$, and our aim is to test $H_0$ against general alternatives.

The problem has been considered in the literature, for a good overview of the existing methods, see e.g. \cite{15,16,01}.
Different approaches to construct a test of CSR include quadrat counts, distance methods (e.g. nearest neighbor and empty spaces), second-order characteristics like the $K$- or the $L$-function, or other measures of dependence. The related problem of testing for uniformity of point patterns with a fixed number of points (i.e. if $\P(N_\lambda=n)=1$ for some $n$)
has been extensively investigated in the univariate case (see \cite{12} for a survey), but also in the multivariate setting, see \cite{06,05,26,02,04,45}.

Our novel idea to test for CSR  is to convert the point data within the observation window $S$ into a binary image,
and then to evaluate this random image by means of the so-called Minkowski functionals from integral geometry, see \cite{19}. Such functionals encompass standard geometric parameters, such as volume, (surface) area, perimeter, and the Euler characteristic, which are robust and efficient shape descriptors that have already been successfully applied to a variety of applications, see \cite{43,44} and the references therein. These data driven and hence random functionals are examined under the assumption of an homogeneous PPP. We determine the mean values and their variance-covariance structure, which opens the ground for different test statistics. Moreover, we analyze an experimental data set by the {\em Fermi Gamma-ray Space Telescope}, see \cite{AceroEtAl2015}. The Fermi sky map includes features whose physical causes are still unknown. New statistical methods could help to clarify some of these open questions.

The paper is organized as follows. In the next section we explain the transition of the data to a binary image using a binning and a threshold procedure. Section \ref{sec:Minkowski.localdependency} deals with calculating the Minkowski functionals in an efficient way, while Section \ref{sec:mean.variance} provides the mean values and the variance structure under the hypothesis. In Section \ref{sec:testH0asy} we propose different test statistics and derive their $H_0$-asymptotics under some suitable limiting regime. The complete covariance structure between the functionals as well as statistics using more than one functional are presented in Section \ref{sec:3Mink.test}. Section \ref{sec:alternativ} is devoted to questions of the behaviour of the tests under an inhomogeneous Poisson Process alternative. In Section \ref{sec:simul} and Section \ref{sec:gamma-ray}, simulation results as well as a data example illustrate the efficiency of the presented methods. We finally state possible extensions and open problems in Section \ref{sec:comm.conc}.



\section{Transition to a binary digital image}\label{sec:Transition.binaryimage}
\begin{figure}[t]
\centering
\includegraphics[scale=0.9]{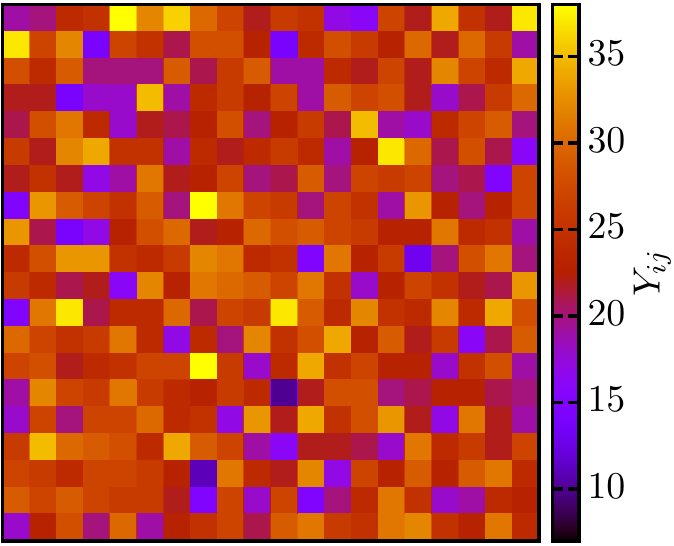}\hspace{0.5cm}\includegraphics[scale=0.9]{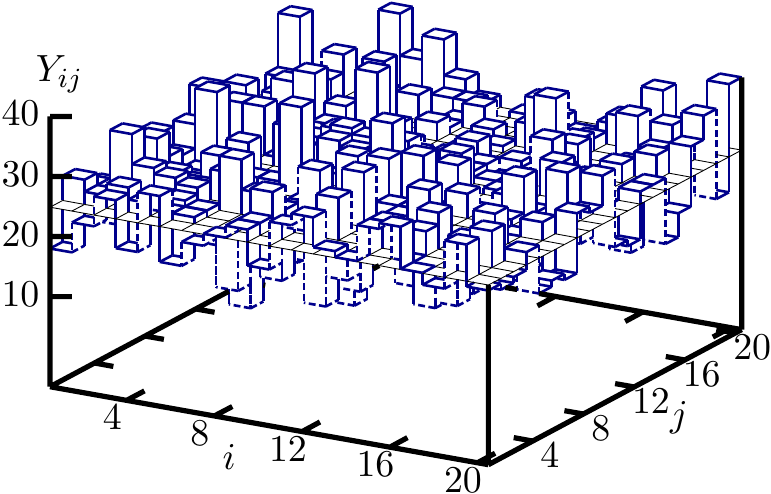}
\caption{Counts map of a realization of a homogeneous Poisson process (left) and visualization of the threshold procedure (right)\label{fig:cm.img}}
\end{figure}

We consider bivariate random point data $X_1,\ldots,X_{N_\lambda}$ in a square observation window that without loss of generality
is taken to be the unit square $S=[0,1]^2$. In a first step, we divide $[0,1]^2$ into $m^2$ pairwise disjoint squares
\begin{equation}\label{cells}
C_{i,j}^{(m)}:=\left[\frac{i-1}{m},\frac{i}{m}\right\ulrcorner\times \left[\frac{j-1}{m},\frac{j}{m}\right\ulrcorner,\quad i,j\in\{1,\ldots,m\},
\end{equation}
that are termed {\em cells} or {\em bins}. Here, the symbol "$\ulrcorner$" stands for a closing round bracket if $i<m$ and/or $j<m$ and for a closing squared bracket if $i=m$ and/or $j=m$. Here and in the sequel we assume $m\ge3$.
We thus have $\bigcup_{i,j=1}^mC_{i,j}^{(m)}=[0,1]^2$.
Denote by $\mathbf{1}\{ A\}$ the indicator function of an event $A$, and let $Y=(Y_{i,j})_{1\le i,j\le m}$ be the random ($m\times m$)-matrix having entries
\begin{equation}
Y_{i,j}=Y_{i,j}^{(m)}:=\sum_{\ell=1}^{N_\lambda}\mathbf{1}\{X_\ell\in C_{i,j}^{(m)}\}.
\end{equation}
Realizations of $Y$ can be visualized by a {\it counts map}, as seen in Figure \ref{fig:cm.img} (left).

\begin{figure}[t]
\centering

\includegraphics{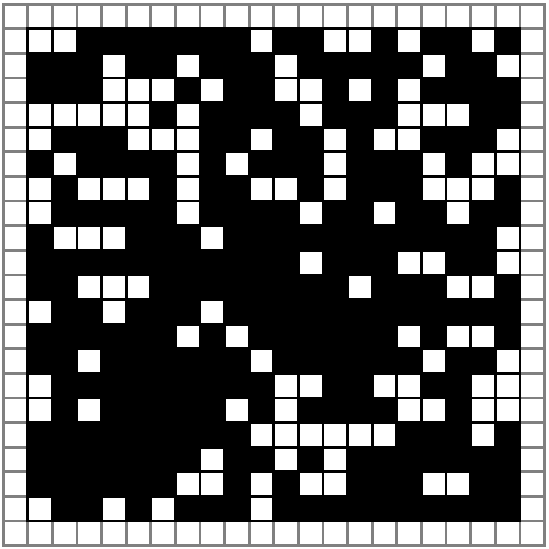}\hspace{1cm}\includegraphics{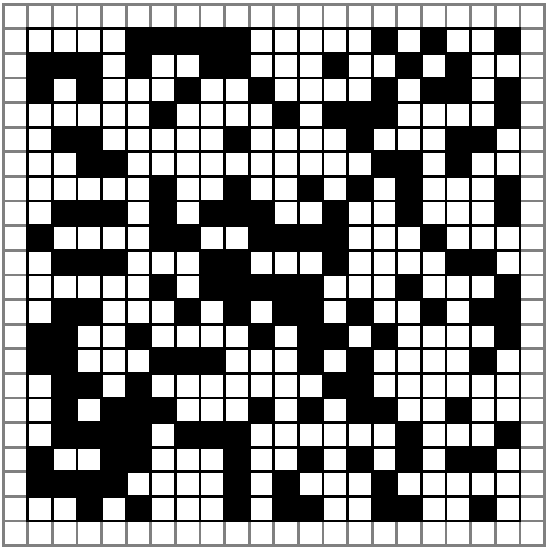}
\caption{Two binary images of an underlying homogeneous Poisson process for different threshold parameters $c$ (left $c=23$, right $c=27$) and fixed cell number parameter $m=20$\label{fig:bin.img}}
\end{figure}

In dependence of a threshold parameter $c\in\N$, we introduce a random $(m+2)\times (m+2)$-matrix $Z=(Z_{i,j})_{i,j=0,\ldots,m+1}$, which is called the {\em digital} or {\em binary image}. Here,
\begin{equation*}
Z_{i,j}=Z_{i,j}^{(m)}(c):=\mathbf{1}\{Y_{i,j}^{(m)}\ge c\}\quad\mbox{if } i,j\in\{1,\ldots,m\}
\end{equation*}
and $Z_{0,j}=Z_{m+1,j}=Z_{i,0}=Z_{i,m+1}=0$ for $i,j=1,\ldots,m$. If we color a cell $C_{i,j}^{(m)}$ black or white according to whether $Z_{i,j}=1$ or $Z_{i,j}=0$,
we obtain a binary (black and white) image, as given in Figure \ref{fig:bin.img}. Notice that, by definition, there is a white border around the cells $C_{i,j}^{(m)}$, $1\le i,j\le m$,
 which is needed for the sake of comparability of the Minkowski functionals. The concept of binary images has wide applications in computer science and image analysis, see \cite{39,38}. In this respect, many algorithmic tools have been developed which are useful in simulations, see for instance \cite{13}.



\section{Minkowski Functionals and Local Dependency}\label{sec:Minkowski.localdependency}
The main idea underlying the new tests of CSR is to evaluate the resulting binary image by means of geometric functionals. In the bivariate case, there are three Minkowski functionals, namely the area, the perimeter, and the Euler characteristic. Knowledge of the values of these functionals does not characterize the binary image.
Nevertheless, Hadwiger's characterization theorem states that every functional acting on nonempty compact convex subsets of $\R^d$, $d\in\N$, that is additive, continuous and invariant under rigid motions can be written as a linear combination of the Minkowski functionals, for details see \cite{19}, p. 628.

A natural question is: Given a $m\times m$ binary image with white border, how can these functionals be computed in an efficient way?
Obviously, we can calculate the area by simply counting the black cells, but the answer for the other functionals is more involved.
However, results from image analysis allow to establish a look-up table like Table \ref{tab:MinkowskiFunctionalsLookUpTable} (even in higher dimensions), see \cite{37,46,47} for early versions in two dimensions and \cite{10} for a survey.
\begin{table}[t]
  \caption{Look-up table for Minkowski functionals: the functional values of the area $A$, the perimeter $P$, and the Euler characteristic $\chi$ are assigned to each $2\times 2$-combination of pixels. The unit of length is the edge-length of a pixel.}
      \label{tab:MinkowskiFunctionalsLookUpTable}
      \vspace*{2mm}
  \centering
  \begin{tabular}{r c     c c c     r c     c c c}
   \toprule
   Configuration &  & $A$ & $P$ & $\chi$ &  Configuration &  & $A$ & $P$ & $\chi$ \\
   \cmidrule(rl){1-5}\cmidrule(rl){6-10}
    1  & \includegraphics[height=9pt]{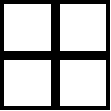}  & 0   & 0 & 0    & 9  & \includegraphics[height=9pt]{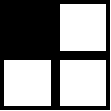}   & 1/4 & 1 & 1/4    \\
    2  & \includegraphics[height=9pt]{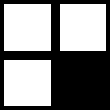}  & 1/4 & 1 & 1/4  & 10 & \includegraphics[height=9pt]{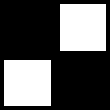}  & 1/2 & 2 & -1/2   \\
    3  & \includegraphics[height=9pt]{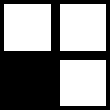}  & 1/4 & 1 & 1/4  & 11 & \includegraphics[height=9pt]{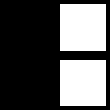}  & 1/2 & 1 & 0     \\
    4  & \includegraphics[height=9pt]{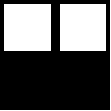}  & 1/2 & 1 & 0    & 12 & \includegraphics[height=9pt]{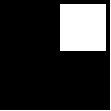}  & 3/4 & 1 & -1/4   \\
    5  & \includegraphics[height=9pt]{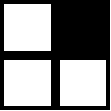}  & 1/4 & 1 & 1/4  & 13 & \includegraphics[height=9pt]{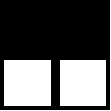}  & 1/2 & 1 & 0      \\
    6  & \includegraphics[height=9pt]{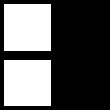}  & 1/2 & 1 & 0    & 14 & \includegraphics[height=9pt]{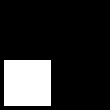}  & 3/4 & 1 & -1/4   \\
    7  & \includegraphics[height=9pt]{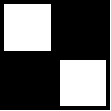}  & 1/2 & 2 & -1/2 & 15 & \includegraphics[height=9pt]{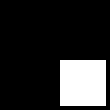}  & 3/4 & 1 & -1/4   \\
    8  & \includegraphics[height=9pt]{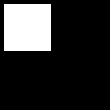}  & 3/4 & 1 & -1/4 & 16 & \includegraphics[height=9pt]{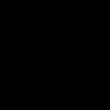}  & 1   & 0 & 0        \\
  \end{tabular}
\end{table}
Here, we move a small $(2\times 2)$-window over the whole digital image (from top left to bottom right) and sum up the values given in the table according to the observed configuration. A basic feature of the structure underlying the look-up table is that each cell can only have an effect on the 8 neighboring cells, which implies a local dependency structure. Given a random digital image $Z$, we thus define
\begin{equation}\label{defamc}
A_{m,c}:=\frac{1}{m}\sum_{i,j=1}^mZ_{i,j}
\end{equation}
as the (scaled) total area covered by 'non-border' cells, the counts of which exceed the threshold $c$, and the perimeter
\begin{equation}\label{defpmc2}
P_{m,c}:=\frac{1}{m}\sum_{i,j=1}^m\psi(Z_{i,j}),
\end{equation}
where, for $i,j\in\{1,\ldots,m\}$,
\begin{equation*}
\psi(Z_{i,j}):=\left\{\begin{array}{cc} 0, & \mbox{if}\;Z_{i,j}=0,\\ 4-(Z_{i-1,j}+Z_{i+1,j}+Z_{i,j-1}+Z_{i,j+1}), & \mbox{if}\;Z_{i,j}=1.\end{array}\right.
\end{equation*}
Motivated by look-up Table \ref{tab:MinkowskiFunctionalsLookUpTable}, we finally study the (scaled) Euler characteristic
\begin{equation}\label{defchimc}
\chi_{m,c}:= \frac{1}{m} \sum_{i,j=1}^{m}W_{i,j},
\end{equation}
where, putting $S_{i,j} := Z_{i,j} + Z_{i+1,j} + Z_{i,j+1}+ Z_{i+1,j+1}$,
\begin{equation} \label{defwij}
W_{i,j} := \left\{\begin{array}{ll}
1/4, & \text{ if } S_{i,j} = 1,\\
-1/4, & \text{ if } S_{i,j} = 3,\\
-1/2, & \text{ if } S_{i,j} = 2 \text{ and }  Z_{i,j}Z_{i+1,j+1} = 1 \text{ or } Z_{i+1,j}Z_{i,j+1} = 1,\\
0, & \text{ otherwise }.
\end{array}\right.
\end{equation}
Each of these functionals is a sum of random variables that depend on the color of the cells $C_{i,j}^{(m)}$. Notice that the summands figuring in (\ref{defpmc2}) and (\ref{defchimc}) are neither independent nor identically distributed.

In view of the $(2\times2)$ observation window that is moved over the binary picture, one can compute each functional as a weighted sum of indicators. As an example, the product
\begin{equation*}
(1-Z_{i,j})(1-Z_{i,j+1})(1-Z_{i+1,j})Z_{i+1,j+1}
\end{equation*}
 is the indicator of the occurrence of configuration 2 of the look-up Table \ref{tab:MinkowskiFunctionalsLookUpTable} at (top left bin) position $(i,j)$. Implementing the look-up table and summing up over all positions with the appropriate weights, one obtains an alternative representation of each of the functionals. For instance, the (scaled) Euler characteristic takes the form
\begin{eqnarray}
\chi_{m,c}&=&\frac1m\sum_{i,j=1}^m\frac14\left(Z_{i,j}^{(m)}+Z_{i,j+1}^{(m)}+Z_{i+1,j}^{(m)}+Z_{i+1,j+1}^{(m)}\right)\nonumber\\
&&-\frac12\left(Z_{i,j}^{(m)}Z_{i+1,j}^{(m)}+Z_{i,j}^{(m)}Z_{i,j+1}^{(m)}+Z_{i,j+1}^{(m)}Z_{i+1,j+1}^{(m)}+Z_{i+1,j}^{(m)}Z_{i+1,j+1}^{(m)}\right.\nonumber\\
&&\left.+2Z_{i,j}^{(m)}Z_{i+1,j+1}^{(m)}+2Z_{i,j+1}^{(m)}Z_{i+1,j}^{(m)}\right)+Z_{i,j}^{(m)}Z_{i,j+1}^{(m)}Z_{i+1,j}^{(m)}\nonumber\\
&&+Z_{i,j}^{(m)}Z_{i,j+1}^{(m)}Z_{i+1,j+1}^{(m)}+Z_{i,j}^{(m)}Z_{i+1,j}^{(m)}Z_{i+1,j+1}^{(m)}+Z_{i,j+1}^{(m)}Z_{i+1,j}^{(m)}Z_{i+1,j+1}^{(m)}\nonumber\\
&&-Z_{i,j}^{(m)}Z_{i,j+1}^{(m)}Z_{i+1,j}^{(m)}Z_{i+1,j+1}^{(m)}.\label{altrepeuler}
\end{eqnarray}
This representation will be used in the proof of Theorem \ref{limitchim}.


\section{Mean and Variance under $H_0$}\label{sec:mean.variance}
To establish a test statistic for CSR we need some characteristics like the mean, the variance and the covariance of the Minkowski functionals under the hypothesis $H_0$.
In case of CSR the random variable $Z_{i,j}$ follows the binomial distribution $\mbox{Bin}(1,p_c)$ where $p_c=\P(Z_{i,j}=1)=\P(Y_{i,j}\ge c)$, depends
on the underlying known intensity $\lambda$ of ${\cal P}_\lambda$ and the threshold $c$. Under $H_0$, we have
\begin{equation}\label{pc}
p_c:=p_c(\lambda,m)=\sum_{k=c}^\infty\exp\left(-\frac{\lambda}{m^2}\right)\frac{(\lambda/m^2)^k}{k!}=1-\exp \left(-\frac{\lambda}{m^2}\right)\sum_{k=0}^{c-1}\frac{(\lambda/m^2)^k}{k!}.
\end{equation}
Clearly, if the parameter $\lambda$ is unknown, we will have to estimate it with a good estimator.
The natural way to estimate the intensity of a stationary planar Poisson process is to count the number of points falling into an observation window,
and to divide this number by the area of the window which, in our case, is one. For more information on estimation techniques see for instance \cite{14}, section 5.5.2.1, or for newer developments \cite{11} and the references therein. By the extreme independence property of homogeneous Poisson processes, the $Z_{i,j}$, $i,j=1,\ldots,m$, are independent, and it follows that
\begin{equation*}
\sum_{i,j=1}^mZ_{i,j}\edist \mbox{Bin}(m^2,p_c),
\end{equation*}
where "$\edist$" means equality in distribution.
Thus, the expected (scaled) total area covered by non-border cells is given by
\begin{equation*}
\mu_A(p_c):=\mathbb{E}\left[A_{m,c}\right]=\frac{1}{m}\sum_{i,j=1}^m\mathbb{E}\left[Z_{i,j}\right]=mp_c.
\end{equation*}
Taking into account the boundary effects, we have
\begin{equation*}
\mathbb{E}\left[\psi(Z_{i,j})\right]=\left\{\begin{array}{cl}
p_c(4-4p_c), \ & \mbox{if}\;i,j\in\{2,\ldots,m-1\},\\
p_c(4-3p_c),\ & \mbox{if}\;i\in\{1,m\},j\in\{2,\ldots,m-1\},\\
p_c(4-3p_c),\ & \mbox{if}\;j\in\{1,m\},i\in\{2,\ldots,m-1\},\\
p_c(4-2p_c),\ & \mbox{if}\;i,j\in\{1,m\},\end{array}\right.
\end{equation*}
and the expected perimeter is given by
\begin{equation*}
\mu_P(p_c):=\mathbb{E}\left[P_{m,c}\right]=\frac{1}{m}\sum_{i,j=1}^m\mathbb{E}\left[\psi(Z_{i,j})\right]=4p_c(m-(m-1)p_c).
\end{equation*}
Moreover, the expected Euler characteristic is
\begin{eqnarray*}
\mu_\chi(p_c)&:=&\mathbb{E}\left[\chi_{m,c}\right]=\frac{1}{m}\sum_{i,j=1}^{m}\mathbb{E}\left[W_{ij}\right]\\
& = & \frac{1}{m} \left(p_c+2(m-1)p_c(1-p_c) +(m-1)^2p_c(1-p_c)(p_c^2-3p_c+1)\right).
\end{eqnarray*}
Thus, each of the mean values of the Minkowski functionals is a function of the probability $p_c$, which in turns depends on $m$, the intensity $\lambda$, and the threshold parameter $c$. The variances are given in the next theorem.
\begin{theorem}\label{varMF}
Under $H_0$ we have
\begin{eqnarray*}
\sigma^2_A:=\BV(A_{m,c})&=&p_c(1-p_c),\\
\sigma^2_{P}:=\BV(P_{m,c})&=&\frac{8}{m^2}p_c(1-p_c)\left((7m^2-13m+4)p_c^2-7m(m-1)p_c+2m^2\right),\\
\sigma^2_\chi:=\BV(\chi_{m,c})&=&\frac{1}{m^2}p_c(1-p_c)\Big{\{}\left(9m^2-30m+25\right)p_c^6-\left(59m^2-194m+159\right)p_c^5\\
&& +\left(137m^2-434m+341\right)p_c^4-\left(139m^2-406m+291\right)p_c^3\\
&&+\left(64m^2-158m+94\right)p_c^2-\left(12m^2-18m+6\right)p_c+m^2\Big{\}}.
\end{eqnarray*}
\end{theorem}
{\bf Proof:} Since the method of computation is the same for each of these formulas, we only illustrate the reasoning by computing $\sigma^2_{P}$.
To this end, observe that the perimeter can alternatively be written as
\begin{equation}\label{defpmc}
P_{m,c}=\frac{1}{m}\sum_{i,j=1}^{m}V_{i,j},
\end{equation}
where, putting $S_{i,j} := Z_{i,j} + Z_{i+1,j} + Z_{i,j+1}+ Z_{i+1,j+1}$ and invoking Table \ref{tab:MinkowskiFunctionalsLookUpTable},
\begin{equation} \label{defvij}
V_{i,j} := \left\{\begin{array}{ll}
1, & \text{ if } S_{i,j} \in \{1,3\},\\
2, & \text{ if } S_{i,j} = 2 \text{ and }  Z_{i,j}Z_{i+1,j+1} = 1 \text{ or } Z_{i+1,j}Z_{i,j+1} = 1,\\
0, & \text{ otherwise}.
\end{array}\right.
\end{equation}
Notice that $V_{i,j}$ assigns one of the values $0,1$ or $2$ to each $(2\times 2)$-window, where the upper left bin is at position $(i,j)$ in the binary picture. We have
\begin{equation*}
\BV(P_{m,c})=\frac{1}{m^2}\left(\sum_{i,j=1}^{m}\BV(V_{i,j})+\sum_{\tiny\begin{array}{c}i,j,k,\ell=1\\(i,j)\not=(k,\ell)\end{array}}^{m+1}\mbox{Cov}(V_{i,j},V_{k,\ell})\right).
\end{equation*}
To compute the sum of the variances, observe that
\begin{eqnarray*}
\BV(V_{1,1})&=&p_c(1-p_c),\\
\BV(V_{1,2})&=&p_c(2-p_c)(1-p_c(2-p_c)),\\
\BV(V_{2,2})&=&4p_c(1-p_c)(3p_c^2-3p_c+1).
\end{eqnarray*}
By symmetry, we have
\begin{equation*}
\sum_{i,j=1}^{m}\BV(V_{i,j})=4\BV(V_{1,1})+4(m-1)\BV(V_{1,2})+(m-1)^2\BV(V_{2,2}).
\end{equation*}
The computation of the sum of covariances uses the methods presented in the proof of Theorem \ref{covarMF}. Summing everything up and simplifying the results leads to the stated formulas.\hfill$\square$



\section{Testing procedures and $H_0$-asymptotics}\label{sec:testH0asy}
In view of the previous sections a natural way to define the new tests is to standardize the data driven Minkowski functionals under $H_0$ in dependence of a fixed threshold parameter $c$. We thus propose
\begin{eqnarray}
  T_A(c;X_1,\ldots,X_{N_\lambda}) &=& \frac1{\sigma_A}\left(A_{m,c}-\mu_A(p_c)\right)^2, \\
  T_P(c;X_1,\ldots,X_{N_\lambda}) &=& \frac1{\sigma_P}\left(P_{m,c}-\mu_P(p_c)\right)^2, \\
  T_\chi(c;X_1,\ldots,X_{N_\lambda}) &=& \frac1{\sigma_\chi}\left(\chi_{m,c}-\mu_\chi(p_c)\right)^2.
\end{eqnarray}
Observe that the variances given in Theorem \ref{varMF} and the probability $p_c$ depend on $c$, $m$ and $\lambda$.
Rejection of $H_0$ is for large values of $T_A, T_P$ or $T_\chi$. For the sake of simplicity we assume that $\lambda$ is known, perhaps on the basis of previous experiments.
 If $\lambda$ has to be estimated, the method of estimation will have effects on the asymptotic statements derived below, as pointed out in \cite{21}.
Throughout this section we assume that $H_0$ holds. To derive the limits in distribution of the Minkowski functionals we consider the limiting regime
\begin{equation}\label{limreg}
\lambda\rightarrow\infty,\,m\rightarrow\infty,\,\frac{\lambda}{m^2}\rightarrow\kappa
\end{equation}
for some $\kappa\in(0,\infty)$. Under this regime, $\lim\BE\left[Y_{i,j}\right]=\kappa$ for each pair $(i,j)$ and the probability $p_c=p_c(\lambda,m)$ figuring in (\ref{pc}) converges to
\begin{equation}\label{pckappa}
p_c(\kappa)=1-\textrm{e}^{-\kappa}\sum_{k=0}^{c-1}\frac{\kappa^k}{k!},
\end{equation}
where $0<p_c(\kappa)<1$. By the central limit theorem we obviously have
\begin{equation*}
\mathcal{A}_{m,\lambda}(c):=\frac{1}{\sigma_A}\left(A_{m,c}-\mu_A(p_c)\right)\cd \mbox{N}(0,1)
\end{equation*}
under (\ref{limreg}), where the symbol $\cd$ means convergence in distribution of random variables and vectors. If $c=1$ and $m$ is fixed, the test based on $A_{m,1}$ is related to the empty boxes test, see \cite{29} if $\P(N_\lambda= n)=1$. Notice that, by the multivariate central limit theorem, we have the convergence in distribution of $(\mathcal{A}_{m,\lambda}(c_1),\ldots, \mathcal{A}_{m,\lambda}(c_s))'$ to some centred $s$-variate normal distribution, for any choice of
$s \ge 2$ and $c_1,\ldots, c_s >0$.  Thus, in any conceivable space of random sequences, there is convergence of finite-dimensional distributions of a random element
${\cal A}_{m,\lambda}$. However, at least for the separable Banach space of sequences converging to zero, equipped with the supremum norm, the double sequence $({\cal A}_{m,\lambda})_{m,\lambda \ge 1}$, where the limit is taken in respect of the limiting regime (\ref{limreg}), is not tight.

Because of the local geometric dependence of the random variables defining the perimeter and the Euler characteristic, we use tools from random geometric graphs, as stated in \cite{09}.
 Let $(I,E)$ be a graph with finite or countable vertex set $I$. For $i,j \in I$, write $i \sim  j$ if $\{i,j\}\in E$, where $E$ is the set of edges.
 For $i \in I$, let ${\cal N}_i:= \{i\} \cup \{j \in I: j \sim i\}$ be the so-called {\em adjacency neighbourhood} of $i$. The graph $(I, \sim )$ is called an {\em dependency graph} for a collection of random variables $(\xi_i, i \in I)$, if for any disjoint subsets $I_1,I_2$ of $I$ such that there are no edges connecting $I_1$ and $I_2$, the collection of random variables $(\xi_i, i \in I_1)$ is independent of $(\xi_i, i \in I_2)$. The following result (\cite{09}, Theorem 2.4), plays a central role in proving the next two statements.
\begin{propos}\label{PRT}
Suppose $(\xi_i)_{i \in I}$ is a finite collection of random variables with dependency graph $(I, \sim )$ having maximum degree $D-1$, where $\BE(\xi_i) =0$ for each $i$. Set $W:= \sum_{i \in I}\xi_i$, and suppose $\BE(W^2) =1$. Then
$$
\sup_{t \in \R}|\mathbb{P}(W \le t) - \Phi(t)| \le \frac{2}{(2 \pi)^{1/4}} \sqrt{ D^2 \sum_{i \in I} \mathbb{E} |\xi_i|^3} + 6 \sqrt{D^3  \sum_{i \in I} \mathbb{E} |\xi_i|^4},
$$
where $\Phi$ is the distribution function of a standard normal distribution {\rm N}$(0,1)$.
\end{propos}
The next result concerns the perimeter $P_{m,c}$.
\begin{theorem}\label{H0P}
For each fixed $c\in\N$ we have under the limiting regime (\ref{limreg})
\begin{equation*}
\frac{1}{\sigma_P}\left(P_{m,c}-\mu_P(p_c)\right)\cd {\rm{N}}(0,1).
\end{equation*}
\end{theorem}
{\bf Proof:}
In view of Theorem \ref{PRT} we choose the vertex set
$$
I := \{i :=(i_1,i_2,i_1-1,i_1+1,i_2-1,i_2+1): (i_1,i_2) \in \{1,\ldots,m\}^2\}
$$
and define
\begin{equation}\label{gl1}
\xi_i := \frac{1}{m\sigma_P}\left(\psi(Z_{i_1,i_2})-\mathbb{E}\left[\psi(Z_{i_1,i_2})\right]\right),\quad i\in I.
\end{equation}
Then $\mathbb{E}(\xi_i)=0$ and $\mathbb{E}\left|\psi(Z_{i_1,i_2})-\mathbb{E}\left[\psi(Z_{i_1,i_2})\right]\right|^\ell<\infty$ if $\ell\in\{3,4\}$ and thus
\begin{equation*}
\mathbb{E}|\xi_i|^\ell=(m\sigma_P)^{-\ell}O(1).
\end{equation*}
To construct a dependency graph we write $i \sim j  : \Longleftrightarrow  |j_1-i_1| + |j_2-i_2| \le 2$ if $i\in I$ is as above and $j :=(j_1,j_2,j_1-1,j_1+1,j_2-1,j_2+1) \in I$.
 Notice that ${\cal N}_i$ has at most 13 elements, which shows that, in our case, the constant $D$ figuring in the statement of Theorem \ref{PRT} is $13$.  With this notation we have
\begin{equation*}
W:=\sum_{i\in I} \xi_i=\frac{1}{\sigma_P}\left(P_{m,c}-\mathbb{E}(P_{m,c})\right).
\end{equation*}
Therefore, since $\sigma_P^2\rightarrow8p_c(\kappa)(1-p_c(\kappa))\left(7p_c(\kappa)^2-7p_c(\kappa)+2\right)$ under the limiting regime (\ref{limreg}), putting $D=13$ and invoking Proposition \ref{PRT} yields
\begin{eqnarray*}
\sup_{t \in \R}|\mathbb{P}(W \le t) - \Phi(t)| &\le& \frac{2}{(2 \pi)^{1/4}} \sqrt{ D^2 \sum_{i \in I} \mathbb{E} |\xi_i|^3} + 6 \sqrt{D^3  \sum_{i \in I} \mathbb{E} |\xi_i|^4}\\
&=&\frac{2D}{(2 \pi)^{1/4}\sqrt{m\sigma_P^3}}O(1)  + \frac{6\sqrt{D^3}}{m\sigma_P^2}O(1)\rightarrow0.
\end{eqnarray*}
\hfill$\square$\\[2mm]
To handle the Euler characteristic, take
$$
I := \{i = (i_1,i_2,i_1+1,i_{2}+1): (i_1,i_2) \in \{1,\ldots,m\}^2\}
$$
and for $i \in I$, put $S_i := Z_{i_1,i_2} + Z_{i_1,i_2+1} + Z_{i_1+1,i_2} + Z_{i_1+1,i_2+1}$.
In this case, with $i$ as above and $j =  (j_1,j_2,j_1+1,j_{2}+1) \in I$, we construct  a dependency graph via
$$
i \sim j : \Longleftrightarrow \max(|i_1-j_1|,|i_2-j_2|) \le 1.
$$
\begin{theorem}
Under the limiting regime (\ref{limreg}), we have
\begin{equation*}
\frac{1}{\sigma_\chi}\left(\chi_{m,c}-\mu_\chi(p_c)\right)\cd \mbox{\rm N}(0,1)
\end{equation*}
for each fixed $c\in\N$.
\end{theorem}
{\bf Proof:} With the dependency graph given above, the proof parallels that of Theorem \ref{H0P}, upon noting that, with
\begin{equation*}
\xi_i^*:=\frac{W_i-\mathbb{E}\left[W_i\right]}{m\sigma_\chi},
\end{equation*}
we have $\mathbb{E}\left[|\xi_i^*|^\ell\right] = O((m\sigma_\chi)^{-\ell})$.\hfill$\square$

The continuous mapping theorem now yields the following result.
\begin{coro}
For fixed $c\in\N$, we have under the limiting regime (\ref{limreg})
\begin{equation*}
    T_j(c;X_1,\ldots,X_{N_\lambda}) \cd \chi^2_1,\quad j\in\{A,P,\chi\}.
\end{equation*}
\end{coro}



\section{Combinations of more than one functional}\label{sec:3Mink.test}
On the basis of promising results regarding the power of tests of $H_0$ against specific alternatives (see Section \ref{sec:simul}), we also considered test statistics
that make use of more than one of the Minkowski functionals. Such an approach requires knowledge of the covariances
$\sigma_{A,P}:=\mbox{Cov}(A_{m,c},P_{m,c})$, $\sigma_{A,\chi}:=\mbox{Cov}(A_{m,c},\chi_{m,c})$ and $\sigma_{P,\chi}:=\mbox{Cov}(P_{m,c},\chi_{m,c})$. These are given as follows.
\begin{theorem}\label{covarMF} Under $H_0$ we have
\begin{eqnarray*}
\sigma_{A,P}&=&\frac{1}{m^2}p_c(1-p_c)(4m^2(1-2p_c)+8mp_c),\\
\sigma_{A,\chi}&=&\frac{1}{m^2}p_c(1\! -\! p_c)\left(-4(m\! -\! 1)^2p_c^3+12(m\! -\! 1)^2p_c^2-4(m\! -\! 1)(2m\! -\! 1)p_c+m^2\right),\\
\sigma_{P,\chi}&=&\frac{4}{m^2}p_c(1-p_c)\Big{\{}\left(6m^2-16m+10\right)p_c^4-\left(22m^2-56m+34\right)p_c^3\\
&& +\left(23m^2-49m+24\right)p_c^2-\left(9m^2-13m+4\right)p_c+m^2\Big{\}}.
\end{eqnarray*}
\end{theorem}
{\bf Proof:} Since the proof is involved due to messy computations, we only show how to compute $\sigma_{P,\chi}$ for the case $m\ge3$. The other covariances are tackled in a similar fashion.
From (\ref{defchimc}) and (\ref{defpmc}), we have
\begin{equation*}
\sigma_{P,\chi}=\frac1{m^2}\sum_{i,j,k,\ell=1}^m\mbox{Cov}(V_{i,j},W_{k,\ell}),
\end{equation*}
where $V_{i,j}$ and $W_{i,j}$ are given in (\ref{defvij}) and (\ref{defwij}), respectively. Notice that,
due to the underlying local dependence structure, the covariance $\mbox{Cov}(V_{i,j},W_{k,\ell})$ vanishes for each pair $(i,j)$ and $(k,\ell)$ of cells that are not neighbors
in the sense that at least one bin of the respective $(2\times 2)$-windows overlaps. For neighboring cells, the resulting covariance depends on how the two cells overlap, giving rise to different 'covariance configurations'. To compute the covariances we have to address the following questions.
\begin{itemize}
\item How many different types of covariance configurations appear in the sum?

\item What are the formulae for the different covariance configurations?

\item How often do we have to count each covariance configuration?
\end{itemize}

As for the first question, observe that, due to the white border of the observation window (see Figure \ref{fig:bin.img}) and the presence of neighboring cells that have joint bins with the border, we have to distinguish the 7 cases 'corners', 'side-corners', 'borders', 'inner-corners', 'inner-side-corners', 'inner-borders' and 'middle cells': Each of theses cases gives rise to a separate covariance configuration. In the same order, the answer to the third question for these configurations is $4,8,4(m-3),4,8,4(m-3),(m-3)^2$. As an example, we compute the covariance formula for two special cases, namely 'corner' and 'side-corner'. For the case 'corner', fixing $V_{1,1}$ and invoking a symmetry argument gives
\begin{equation}\label{corncov}
\mbox{Cov}(V_{1,1},W_{1,1})+2\mbox{Cov}(V_{1,1},W_{1,2})+\mbox{Cov}(V_{1,1},W_{2,2}),
\end{equation}
since the upper left $(2\times2)$-window has four neighboring $(2\times2)$-windows with intersecting bins.
Since, under $H_0$, the colorings of the single bins are independent, the summands above read
\begin{equation*}
\mbox{Cov}(V_{1,1},W_{1,1})=\frac14p_c(1-p_c), \quad \mbox{Cov}(V_{1,1},W_{1,2})=\frac14p_c(1-p_c)(1-2p_c),
\end{equation*}
and
\begin{equation*}
\mbox{Cov}(V_{1,1},W_{2,2})=-\frac14p_c(1-p_c)(4p_c^3-12p_c^2+8p_c-1).
\end{equation*}
Thus, the sum figuring in (\ref{corncov}) equals $p_c(1-p_c)^4$, which is the contribution to the total covariance of each of the corners. For the case 'side-corner' we fix $V_{1,2}$ and, again due to symmetry, have to consider five summands, namely
\begin{eqnarray*}
\mbox{Cov}(V_{1,2},W_{1,1})+\mbox{Cov}(V_{1,2},W_{1,2})+2\mbox{Cov}(V_{1,2},W_{1,3})\\
+\mbox{Cov}(V_{1,2},W_{2,2})+\mbox{Cov}(V_{1,2},W_{2,3}).
\end{eqnarray*}
Calculation of each summand and summing up gives the contribution $p_c(1-2p_c)(2-p_c)(1-p_c)^3$ to the total covariance for each 'side-corner'-case of pairs of cells. Counting the number of times that each of the different configurations that yield a non-vanishing contribution to the total covariance can occur and summing up, the final result follows from tedious calculations.\hfill$\square$\\

The formulas figuring in Theorem \ref{covarMF} have been simplified using the CAS Maple 18, and they have been checked by Monte Carlo simulations in {\tt R}. The complete covariance structure between the Minkowski functionals is given by the symmetric ($3\times 3$)-matrix
\begin{equation*}
\Sigma_{c,m,\lambda}:=\left(\begin{array}{ccc}\sigma^2_A & \sigma_{A,P} & \sigma_{A,\chi}\\
\sigma_{A,P} & \sigma^2_{P} & \sigma_{P,\chi}\\
\sigma_{A,\chi} & \sigma_{P,\chi} & \sigma^2_\chi
\end{array}\right).
\end{equation*}
The index stresses the dependence of the covariance structure on the threshold parameter $c$, the underlying intensity $\lambda$ of the PPP, and on $m$. The determinant of $\Sigma_{c,m,\lambda}$ is given by
\begin{eqnarray*}
\det\left(\Sigma_{c,m,\lambda}\right)&=&\frac8{m^6}p_c^5(1-p_c)^3\Big{\{}-(m^2-3m+4)^3p_c^6\\
&&+2(2m^2-5m+6)(m^2-3m+4)(m^2-5m+8)p_c^5\\
&&+(-4m^6+50m^5-280m^4+878m^3-1580m^2+1552m-704)p_c^4\\
&&+(-4m^6+22m^5-302m^3+828m^2-880m+384)p_c^3\\
&&+\left(11m^6-77m^5+229m^4-275m^3+56m^2+112m-64\right)p_c^2\\
&& -4m(m-1)(2m^4-9m^3+20m^2-18m+4)p_c+2m^3(m-1)^3\Big{\}}.
\end{eqnarray*}


According to Maple 18, there is an explicit representation of the inverse of $\Sigma_{c,m,\lambda}$, which shows that
for $0<p_c<1$ the matrix $\Sigma_{c,m,\lambda}$ is nonsingular for each $p_c \in (0,1)$. This expression, however,  is too
complicated to be reproduced here.
Letting  $m\rightarrow\infty$ we obtain the asymptotic covariance matrix
\begin{eqnarray*}
\Sigma&:=&p_c(1-p_c) \left\{\diag(0,0,9)p_c^6+\diag(0,0,-59)p_c^5+\left(\begin{array}{ccc} 0 & 0 & 0\\ 0 & 0 & 24 \\ 0 & 24 & 137\end{array}\right)p_c^4 \right. \\
&& \left. +\left(\begin{array}{ccc} 0 & 0 & -4\\ 0 & 0 & -88 \\ -4 & -88 & -139\end{array}\right)p_c^3
+\left(\begin{array}{ccc} 0 & 0 & 12\\ 0 & 56 & 92 \\ 12 & 92 & 64\end{array}\right)p_c^2
+\left(\begin{array}{ccc} 0 & -8 & -4\\ -8 & -56 & -36 \\ -4 & -36 & -12\end{array}\right)p_c \right. \\
&& \left. +\left(\begin{array}{ccc} 1 & 4 & 1\\ 4 & 16 & 4 \\ 1 & 4 & 1\end{array}\right)\right\}.
\end{eqnarray*}
This is nonsingular if $0<p_c<1$, since the inverse matrix is given by

\begin{eqnarray*}
\Sigma^{-1}&:=&\left(p_c^2(1-p_c)^4(p_c^2-2)\right)^{-1}\cdot \left\{ \diag(-9,0,0)p_c^6 \right.\\
&& \left. +\left(\begin{array}{ccc} 43 & -3 & 0\\ -3 & 0 & 0 \\ 0 & 0 & 0\end{array}\right)p_c^5
+\left(\begin{array}{ccc} -87 & 27/2 & 0\\ 27/2 & -7/8 & 0 \\ 0 & 0 & 0\end{array}\right)p_c^4
+\left(\begin{array}{ccc} 103 & -22 & 4\\ -22 & 15/4 & 0 \\ 4 & 0 & 0\end{array}\right)p_c^3  \right.\\
&& \left. +\left(\begin{array}{ccc} -76 & 35/2 & -8\\ 35/2 & -41/8 & 1 \\ -8 & 1 & 0\end{array}\right)p_c^2
+\left(\begin{array}{ccc} 30 & -8 & 4\\ -8 & 5/2 & -2 \\ 4 & -2 & 0\end{array}\right)p_c
+\left(\begin{array}{ccc} -5 & 3/2 & -1\\ 3/2 & -1/2 & 1/2 \\ -1 & 1/2 & -1\end{array}\right)\right\} .
\end{eqnarray*}

Since the mean, the variance, and the covariance structure have been computed under $H_0$, one may expect that, for each value of the threshold parameter $c$, the standardized vector
\begin{equation*}
\Sigma_{c,m,\lambda}^{-\frac{1}{2}}\left(\left(A_{m,c}, P_{m,c}, \chi_{m,c}\right)^\top-\mu_c\right)
\end{equation*}
does not deviate too much from the origin in $\R^3$. Here, $\Sigma_{c,m,\lambda}^{-\frac{1}{2}}$ denotes the symmetric square root of $\Sigma_{c,m,\lambda}^{-1}$, $\top$ stands
for the transposition of vectors and matrices, and $\mu_c:=(\mu_A,\mu_P,\mu_{\chi})^\top$. Writing $\|\cdot\|$ for the Euclidean norm, we define a family of test statistics depending on $c$, namely
\begin{eqnarray*}
T_c(X_1,\ldots,X_{N_\lambda})& := & \left\|\Sigma_{c,m,\lambda}^{-\frac{1}{2}}\left(\left(A_{m,c}, P_{m,c}, \chi_{m,c}\right)^\top-\mu_c\right)\right\|^2\\
& = & \left(\! \left(A_{m,c}, P_{m,c}, \chi_{m,c}\right)^\top\! -\! \mu_c\right)^\top \Sigma_{c,m,\lambda}^{-1}\left(\! \left(A_{m,c}, P_{m,c}, \chi_{m,c}\right)^\top\! -\! \mu_c\right).
\end{eqnarray*}
An asymptotic equivalent alternative to this statistic is
\begin{equation*}
\widetilde{T}_c(X_1,\ldots,X_{N_\lambda}):=\left(\left(A_{m,c}, P_{m,c}, \chi_{m,c}\right)^\top-\mu_c\right)^\top \Sigma^{-1}\left(\left(A_{m,c}, P_{m,c}, \chi_{m,c}\right)^\top-\mu_c\right).
\end{equation*}
Rejection of $H_0$ is for large values of $T_c$ or $\widetilde{T}_c$.

In what follows, we state the asymptotic distributions of  $T_c$ and $\widetilde{T}_c$ under $H_0$. As in Section \ref{sec:testH0asy} our main problem is the local dependency structure of the vector $(A_{m,c}, P_{m,c}, \chi_{m,c})$.
\begin{theorem}
Under the limiting regime (\ref{limreg}), we have for fixed $c\in\N$
\begin{enumerate}
\item[a)] $T_c\cd\chi^2_3$, \\
\item[b)] $\widetilde{T}_c\cd\chi^2_3$.
\end{enumerate}
\end{theorem}
{\bf Proof:} We use Theorem 2.2 of \cite{20}. To this end, fix $(i,j) \in \{1,\ldots,m\}^2$, and let
$S_{(i,j)}$ be the set of indices of the points that are neighbors of $(i,j)$, enlarged by $\{(i,j)\}$. Moreover, put
$\displaystyle\mathcal{N}_{(i,j)}=\textstyle{\bigcup_{(k,\ell)\in S_{(i,j)}}} S_{(k,\ell)}$. The set
$S_{(i,j)}$ has at most 9 elements, and the cardinality of ${\cal N}_{(i,j)}$ is at most 81.
Arguing as in the example on p. 338 of \cite{20}, we see that each of the constants $\chi_1,\chi_2,\chi_3$ figuring in formula (2.2) of \cite{20} vanishes.
Suppose $\mathcal{H}$ is a class of measurable functions from $\R^3$ to $\R$ which is closed under affine transformation of the argument
 and satisfies the conditions on p. 335 of \cite{20}. Theorem 2.2 of \cite{20} then states that, for constants $\alpha_1,\alpha_2$ and $\boldmath{W}:= (A_{m,c}, P_{m,c}, \chi_{m,c})$, we have
\begin{eqnarray*}
\sup\{|\BE(h(\boldmath{W}))\! - \! \Phi h|:h\! \in \! \mathcal{H}\}\le 81\alpha_1\left(\alpha_2B_m+9m^2\alpha_2B_m^3(|\log B_m|+2\log m)\! \right).
\end{eqnarray*}
Here, $B_m$ is $o(1/m)$,  and $\Phi h=\int_{\R^3}h(z)\Phi(dz)$, where $\Phi$ denotes the multivariate standard normal distribution function.
 From the invariance of affine transformations of $\mathcal{H}$, we therefore have under the limiting regime (\ref{limreg})
\begin{equation*}
\Sigma_{c,m,\lambda}^{-\frac{1}{2}}(W-\BE(W))\cd \textrm{N}_3(0,\textrm{I}_3),
\end{equation*}
where $\textrm{N}_3(0,\textrm{I}_3)$ denotes a centered three-dimensional normal distribution with unit covariance matrix.
Assertion a) then follows from the continuous mapping theorem. Since $\Sigma_{c,m\lambda} \to \Sigma$ under the limiting regime (\ref{limreg}),
assertion b) is a consequence of a) and Slutzky's Lemma. \hfill$\square$.




\section{Asymptotics under alternatives}\label{sec:alternativ}
A feasible alternative could be the following:
Let $f$ be a continuous Lebesgue density over $[0,1]^2$. Suppose ${\cal P}_{\lambda f} := \{X_1,\ldots, X_{N_\lambda}\}$
is a Poisson process on $[0,1]^2$ with intensity function $\lambda f$, i.e., $(X_j)_{j\ge 1}$ is a sequence of i.i.d. random variables with
density $f$ and $N_\lambda \edist \text{Po}(\lambda)$, independent of $(X_j)_{j \ge 1}$. For a Borel subset $A$ of $[0,1]^2$, let ${\cal P}_{\lambda f}(A) := \sum_{i=1}^{N_\lambda} {\bf 1}\{X_i \in A\}$
be the number of points of ${\cal P}_{\lambda f}$ in $A$. Then, putting $p_A := \int_A f(x) \mbox{d} x$, and conditioning on $N_\lambda$,
we have ${\cal P}_{\lambda f}(A) \edist \text{Po}\left(\lambda \int_A f(x) \mbox{d} x\right)$. Moreover, for any pairwise disjoint Borel sets $B_1,\ldots,B_\ell$ of $[0,1]^2$, the random variables ${\cal P}_{\lambda f}(B_1), \ldots, {\cal P}_{\lambda f}(B_\ell)$ are independent. Let $A_{m,c},P_{m,c} $ and $\chi_{m,c}$ be defined as in (\ref{defamc}), (\ref{defpmc2}) and (\ref{defchimc}), respectively, where, for fixed $c \in \N$, $Z^{(m)}_{i,j} := {\bf 1}\{{\cal P}_{\lambda f}(C^{(m)}_{i,j}) \ge c\}$
and $C^{(m)}_{i,j}$ as in (\ref{cells}). Since $f$ is continuous, we have under the limiting regime (\ref{limreg})
\begin{equation}\label{asygl}
a_{i,j}^{(m)}:=\lambda \int_{C^{(m)}_{i,j}} f(x) \mbox{d} x \sim \kappa f\left(\frac{i}{m},\frac{j}{m}\right),
\end{equation}
where $\sim$  means asymptotic equivalence under the limiting regime. Moreover, since $[0,1]^2$ is compact, $f$ is uniformly continuous over $[0,1]^2$ by the Heine--Cantor theorem. Writing an unspecified integral for integration over the unit square, and denoting
\begin{equation}\label{qckapf}
q_{c,\kappa,f}(x):=\sum_{k=0}^{c-1} \frac{\kappa^k}{k!} f(x)^k {\rm{e}}^{-\kappa f(x)},\quad c\ge1,\kappa>0,x\in[0,1]^2,
\end{equation}
we have the following result.
%
%
\begin{theorem}\label{cons.Amc}
Under ${\cal P}_{\lambda f}$ and the limiting regime (\ref{limreg}), we have for fixed $c\in\N$
\begin{equation*}
\frac1mA_{m,c}\cas1 - \int q_{c,\kappa,f}(x) {\rm{d}} x.
\end{equation*}
\end{theorem}
{\bf Proof:}
Invoking (\ref{asygl}) we have
\begin{eqnarray*}
\BE\left[A_{m,c}\right] & = & \frac{1}{m} \sum_{i,j=1}^m \PP \left({\cal P}_{\lambda f}\left(C^{(m)}_{i,j}\right) \ge c \right)\\
& = & \frac{1}{m} \sum_{i,j=1}^m  \left( 1- \sum_{k=0}^{c-1} \exp\left(- a^{(m)}_{i,j} \right) \, \frac{1}{k!} \, \left(a^{(m)}_{i,j} \right)^k \right)\\
& = & m - \sum_{k=0}^{c-1} \frac{1}{k!} \sum_{i,j=1}^m \frac{1}{m} \exp\left(-a^{(m)}_{i,j} \right) \, \left(a^{(m)}_{i,j} \right)^k,
\end{eqnarray*}
and thus, using the asymptotic equivalence in (\ref{asygl}),
\begin{equation}\label{glewert}
\lim \BE\left(\frac1mA_{m,c}\right) = 1 - \sum_{k=0}^{c-1} \frac{\kappa^k}{k!} \int f(x)^k \mbox{e}^{-\kappa f(x)} \mbox{d} x.
\end{equation}
Since $\BV(m^{-1}A_{m,c}) \le m^{-2}$, Tschebyshev's inequality gives
$$
\sum_{m=1}^\infty \PP\left(\frac1m|A_{m,c} - \BE A_{m,c}| \ge \varepsilon \right) < \infty
$$
for each positive $\varepsilon$. The lemma of Borel-Cantelli yields $\frac1m(A_{m,c} - \BE A_{m,c}) \to 0$ $\PP$-a.s. In view of \eqref{glewert}, we are done.\hfill$\square$\\[2mm]

Putting
\[
L_c(u) := \frac{1}{(c-1)!} \int_0^u \mbox{e}^{-t} \, t^{c-1} \, \mbox{d} t, \qquad u >0,
\]
for $c \in \mathbb{N}$, monotone convergence and a well-known relation between probabilities of level exceedances of Poisson distributions and the lower incomplete
Gamma function yield
\begin{eqnarray*}
1 - \int q_{c,\kappa,f}(x) \mbox{d} x & = & \int \sum_{k=c}^\infty \frac{\kappa^k}{k!} f(x)^k \mbox{e}^{-\kappa f(x)} \mbox{d} x\\
& = & \int \PP\left(\text{Po}(\kappa f(x)) \ge c\right) \mbox{d} x\\
& = & \int \frac{1}{(c-1)!} \int_0^{\kappa f(x)} \mbox{e}^{-t} t^{c-1} \mbox{d} t \, \mbox{d} x\\
& = & \int L_c\left(\kappa f(x)\right) \, \mbox{d} x.
\end{eqnarray*}
If $c=1$, Jensen's inequality shows that this expression attains its maximum value $1-\mbox{e}^{-\kappa}$ if, and only if, $f$ is the uniform density  over $[0,1]^2$.
Such a result that characterizes the uniform distribution by an extremal property does no longer hold if $c \ge 2$, since, as is readily seen,
the function $L_c$ is strictly convex on $(0,c-1)$ and strictly concave on $(c,\infty)$. This observation is connected to a two-crossings theorem regarding mixtures from
distributions that belong to exponential families, see \cite{40} or \cite{41}, p. 39.
\begin{theorem}
Under ${\cal P}_{\lambda f}$ and the limiting regime (\ref{limreg}), we have  for fixed $c\in\N$
\begin{equation*}
\frac1mP_{m,c}\cas I_{c,\kappa}(f),
\end{equation*}
where
\begin{equation*}
I_{c,\kappa}(f) := 4 \left(\int q_{c,\kappa,f}(x)\mbox{d} x -\int q_{c,\kappa,f}^2(x) \mbox{d} x \right).
\end{equation*}
\end{theorem}
{\bf Proof:} For the (scaled) perimeter $P_{m,c}$ we have
$$
\BE \left[P_{m,c}\right] = \frac{1}{m} \sum_{i,j=1}^m \BE\left( Z_{i,j}^{(m)} \left[4- Z_{i-1,j}^{(m)} - Z_{i+1,j}^{(m)} - Z_{i,j-1}^{(m)} - Z_{i,j+1}^{(m)}  \right]\right).
$$
By the complete independence property of ${\cal P}_{\lambda f}$,
\begin{eqnarray*}
\BE \left[Z_{i,j}^{(m)} Z_{i-1,j}^{(m)}\right] & = & \PP\left({\cal P}_{\lambda f}\left(C^{(m)}_{i,j}\right) \ge c, {\cal P}_{\lambda f}\left(C^{(m)}_{i-1,j}\right) \ge c \right)\\
& = & \PP\left({\cal P}_{\lambda f}\left(C^{(m)}_{i,j}\right) \ge c\right) \PP\left({\cal P}_{\lambda f}\left(C^{(m)}_{i-1,j}\right) \ge c \right),
\end{eqnarray*}
and likewise for $\BE \left[Z_{i,j}^{(m)} Z_{i+1,j}^{(m)}\right] $ etc. With $a^{(m)}_{i,j}$ in (\ref{asygl}) we have
\begin{equation}\label{Plambdaf}
\PP \left({\cal P}_{\lambda f}\left(C^{(m)}_{i,j}\right) \ge c \right) = 1- \sum_{k=0}^{c-1} \exp\left(-  a^{(m)}_{i,j} \right) \, \frac{1}{k!} \, \left(a^{(m)}_{i,j} \right)^k
\end{equation}
and thus
\begin{eqnarray*}
\BE \left[Z_{i,j}^{(m)} Z_{i-1,j}^{(m)}\right] & = & 1 \! - \! \sum_{k=0}^{c-1} \! \exp \! \left( \! -  a^{(m)}_{i,j} \right) \frac{1}{k!}  \left( \! a^{(m)}_{i,j} \! \right)^k
- \sum_{\ell=0}^{c-1} \! \exp \! \left(\! -  a^{(m)}_{i-1,j}
 \right) \frac{1}{\ell!}  \left(\! a^{(m)}_{i-1,j} \!
 \right)^\ell\\
 & & \qquad + \sum_{k,\ell=0}^{c-1} \frac{1}{k!\ell!} \exp\left(-  a^{(m)}_{i,j} - a^{(m)}_{i-1,j} \right) \left(a^{(m)}_{i,j} \right)^k \left(a^{(m)}_{i-1,j} \right)^\ell .
\end{eqnarray*}
It follows that
\begin{eqnarray*}
\frac{1}{m} \sum_{i,j=1}^m \BE \left[Z_{i,j}^{(m)} Z_{i-1,j}^{(m)}\right] & = & m - \sum_{k=0}^{c-1} \frac{1}{k!} \sum_{i,j=1}^m \frac{1}{m} \exp\left(-  a^{(m)}_{i,j} \right) \left(a^{(m)}_{i,j} \right)^k\\
& & \ - \sum_{\ell=0}^{c-1} \frac{1}{\ell!} \sum_{i,j=1}^m \frac{1}{m} \exp \! \left(\! -  a^{(m)}_{i-1,j} \!  \right) \!  \left( \! a^{(m)}_{i-1,j} \! \right)^\ell\\
 & & \ + \sum_{k,\ell=0}^{c-1} \frac{1}{k!\ell!} \sum_{i,j=1}^m \frac{1}{m} \! \exp\! \left(\! -  a^{(m)}_{i,j}\! - \! a^{(m)}_{i-1,j} \! \right)
 \left(\! a^{(m)}_{i,j} \! \right)^k \! \left(\! a^{(m)}_{i-1,j} \! \right)^\ell .
\end{eqnarray*}
In view of \eqref{asygl}, the uniform continuity of $f$ and a symmetry argument give
\begin{eqnarray*}
\lim \frac{1}{m^2} \sum_{i,j=1}^m \BE \left[Z_{i,j}^{(m)} Z_{i-1,j}^{(m)}\right] & = & 1 - 2 \sum_{k=0}^{c-1} \frac{\kappa^k}{k!} \int \mbox{e}^{-\kappa f(x)} f(x)^k \mbox{d} x\\
& & \qquad + \sum_{k,\ell=0}^{c-1} \frac{\kappa^{k+\ell}}{k!\ell!} \int \textrm{e}^{-2\kappa f(x)} f(x)^{k+\ell} \mbox{d} x.
\end{eqnarray*}
The same limits arise if we consider
$$
\frac{1}{m^2} \sum_{i,j=1}^m \BE \left[Z_{i,j}^{(m)} Z_{i-1,j}^{(m)}\right], \ \frac{1}{m^2} \sum_{i,j=1}^m \BE \left[Z_{i,j}^{(m)} Z_{i,j-1}^{(m)}\right] \mbox{ and }
\frac{1}{m^2} \sum_{i,j=1}^m \BE \left[Z_{i,j}^{(m)} Z_{i,j+1}^{(m)}\right].
$$
Since, by Theorem \ref{cons.Amc}, we have
\[
\frac{1}{m^2} \sum_{i,j=1}^m \BE\left[Z_{i,j}^{(m)}\right] \rightarrow 1- \int  q_{c,\kappa,f}(x)\, \textrm{d} x
\]
almost surely, it follows that $\lim \BE(m^{-1}P_{m,c}) = I_{c,\kappa}(f)$. Since $\BV(m^{-1} P_{m,c}) \le C_1/m^2$ for some finite constant $C_1$, we have $\lim P_{m,c} =I_{c,\kappa}(f)$ almost surely under the limiting regime (\ref{limreg}).\hfill$\square$
\begin{theorem}\label{limitchim}
Under ${\cal P}_{\lambda f}$ and the limiting regime (\ref{limreg}), we have for fixed $c\in\N$
\begin{equation*}
\frac1m\chi_{m,c}\cas J_{c,\kappa}(f),
\end{equation*}
where
\begin{equation*}
J_{c,\kappa}(f):=1-\int  q_{c,\kappa,f}(x)\, \textrm{d} x-2\left(1-\int  q^2_{c,\kappa,f}(x)\, \textrm{d} x\right)+1-\int  q^4_{c,\kappa,f}(x)\, \textrm{d} x.
\end{equation*}
\end{theorem}
{\bf Proof:} In view of the techniques used in the previous proofs and formula (\ref{altrepeuler}), we have to compute
\begin{equation*}
\BE\left[Z_{i,j}^{(m)}Z_{i,j+1}^{(m)}Z_{i+1,j}^{(m)}\right]\quad\mbox{and}\quad \BE\left[Z_{i,j}^{(m)}Z_{i,j+1}^{(m)}Z_{i+1,j}^{(m)}Z_{i+1,j+1}^{(m)}\right].
\end{equation*}
The details are omitted.\hfill$\square$\\[2mm]
Notice that if  $f$ is the uniform density on $[0,1]^2$ then
\begin{equation*}
\int  q_{c,\kappa,f}(x)\, \textrm{d} x=1-p_c(\kappa),\quad \int  q^2_{c,\kappa,f}(x)\, \textrm{d} x=(1-p_c(\kappa))^2
\end{equation*}
and
\begin{equation*}
\int  q^4_{c,\kappa,f}(x)\, \textrm{d} x =(1-p_c(\kappa))^4,
\end{equation*}
where $p_c(\kappa)$ is given in (\ref{pckappa}). It is easily checked that the almost sure limits obtained are consistent with the formulas of the mean values under $H_0$, divided by $m$, with respect to the limiting regime (\ref{limreg}).


\section{Simulations}\label{sec:simul}
In this section we compare the finite-sample power of the test based on a single Minkowski functional, i.e. $T_j$, $j\in\{A,P,\chi\}$, as well as the tests based on $T_c$ and $\widetilde{T}_c$ that make use of all three functionals, with the power of several competitors. All simulations are performed using the statistical computing environment {\tt R}, see \cite{31}. Notice that, strictly speaking, the new procedures form a two-parametric class of tests, depending on the threshold parameter $c$, the mean number (under $H_0$) of points in each bin $\kappa$ and the number $m^2$ of bins. The latter parameter is chosen to fulfill the limiting regime (\ref{limreg}), and throughout this section we fix $m:=m(\lambda,\kappa)=\scriptstyle\left\lfloor\sqrt{\lambda\slash\kappa}\right\rfloor$, where $\lfloor\cdot\rfloor$ is the floor function. Observe that no other choice of $m$ has been considered, so that one might find combinations of $m,c$ and $\kappa$ that result in a better power performance. The intensity $\lambda$ of the simulated processes has to be estimated in a prior independent experiment and is therefore considered to be known. In each scenario we consider the intensities $\lambda\in\{50,100,200,500\}$, and the nominal level of significance is set to $0.05$. Empirical critical values under $H_0$ for $T_j$, $j\in\{A,P,\chi\}$, $T_c$ and $\widetilde{T}_c$ have been simulated with $100~000$ replications (see Tables \ref{tab:quantiles.univ} and \ref{tab:quantiles.multiv}), and each entry in Tables \ref{tab:inh.PPP} and \ref{tab:kappa3} referring to the power of the tests is based on $10~000$ replications.

\begin{table}[t]
\centering
\caption{Empirical $95\%$ quantiles of $T_j$, $j\in\{A,P,\chi\}$, for $\kappa=1$}\label{tab:quantiles.univ}

\begin{tabular}{c|ccc||ccc||ccc}
 test & \multicolumn{3}{c}{$T_A$} & \multicolumn{3}{c}{$T_P$} & \multicolumn{3}{c}{$T_\chi$}\\\hline
$\lambda\backslash c$ & 1 & 2 & 5 & 1 & 2 & 5 & 1 & 2 & 5\\ \hline
 50     & 3.93 & 4.11 & 3.33 & 4.18 & 3.77 & 3.38 & 3.52 & 4.85 & 3.48\\
 100    & 3.65 & 3.78 & 7.32 & 3.89 & 3.69 & 7.42 & 4.41 & 3.21 & 1.16\\
 200    & 3.95 & 3.88 & 1.91 & 3.87 & 3.89 & 1.95 & 3.74 & 4.22 & 2.03\\
 500    & 3.88 & 3.82 & 4.37 & 3.70 & 3.87 & 4.45 & 3.97 & 3.64 & 2.08\\
 1000   & 3.70 & 3.85 & 3.58 & 3.83 & 3.84 & 3.67 & 3.94 & 4.00 & 3.89\\
 10000  & 3.82 & 3.85 & 3.69 & 3.85 & 3.86 & 3.70 & 3.90 & 3.88 & 4.07\\
\hline
\end{tabular}
\end{table}

\begin{table}[t]
\centering
\caption{Empirical $95\%$ quantiles of $T_c$ and $\widetilde{T}_c$ for $\kappa=1$}\label{tab:quantiles.multiv}

\begin{tabular}{c|ccc||ccc}
 test & \multicolumn{3}{c}{$T_c$} & \multicolumn{3}{c}{$\widetilde{T}_c$} \\\hline
$\lambda\backslash c$ & 1 & 2 & 5 & 1 & 2 & 5 \\ \hline
 50     & 7.86 & 8.13 & 3.55  & 30.10 & 9.55 & 3.59\\
 100    & 7.81 & 7.95 & 7.78  & 21.54 & 8.93 & 7.84\\
 200    & 7.84 & 7.88 & 2.10  & 17.38 & 8.47 & 6.88\\
 500    & 7.83 & 7.92 & 4.87  & 13.49 & 8.22 & 4.89\\
 1000   & 7.79 & 7.83 & 26.54 & 11.65 & 8.03 & 25.68\\
 10000  & 7.84 & 7.79 & 11.73 & 9.26  & 7.95 & 8.98\\

\hline
\end{tabular}
\end{table}

Notice that the $95\%$ quantile of $\chi^2_1$ is 3.84, and that of $\chi^2_3$ is 7.81.
The parameter $m$ is chosen in such a way that, under $H_0$, the average number of points that fall into one bin  is one.
 Consequently, the fluctuation of the critical values for $c=5$ may result from a too small number of black bins.

As competitors to the new tests we considered the following procedures, which are all standard methods included in the package {\tt spatstat}. We chose these procedures to have representatives of the different approaches, namely quadrat counts, distance methods, and methods based on the $K$- or the $L$-function.
\begin{itemize}
\item[(i)] For the quadrat count $\chi^2$-test, see \cite{spatstat}, one divides the observation window into disjoint squares $B_1,\ldots,B_k$
with equal area $1/k$ and counts the number of points $U_1,\ldots,U_k$ in each square.
Under $H_0$ the $U_j$ are independent Poisson random variables with expected value $\lambda/k$. Given the total number of points $N_\lambda=\sum_{j=1}^kU_j$ the expected count in square $B_j$ is $N_\lambda/k$. The test statistic is then (see \cite{spatstat}, p. 165, display (6.5))
    \begin{equation*}
    Q=\sum_{j=1}^k\frac{(U_j-N_\lambda/k)^2}{N_\lambda/k}.
    \end{equation*}
    Under the null hypothesis, the limit law of $Q$ is a $\chi^2_{k-1}$ distribution. Notice that one should choose $k$ in order to obtain expected counts greater than 5 in each square.
    Otherwise the approximation of the critical values is too far away from the theoretical quantiles. Hence we chose $\sqrt{k}:=\lfloor \sqrt[4]{\lambda}\rfloor$ to guarantee sufficiently many points in each square.

\item[(ii)] Hopkins and Skellam (see \cite{33,34} and, for more details, \cite{spatstat}, p. 259)
proposed a test based on the combination of nearest neighbor distances and empty space distances. Consider a subsample of size $n$ of the data and compute the nearest neighbor distances $D_i$, $i=1,\ldots,n$, and the empty-space distances $E_j$, $j=1,\ldots,n$ for an equal number $n$ of uniformly sampled spatial locations. Then the Hopkins-Skellam index is given by
    \begin{equation*}
    H=\frac{\sum_{i=1}^nD_i}{\sum_{j=1}^nE_j}.
    \end{equation*}
      Under the null hypothesis $H$ is distributed according to an $F_{2n,2n}$-distribution. As remarked in \cite{35} one should choose
      $n\le N_\lambda/10$ since the distributional theory is only known for a sparsely sampled homogeneous Poisson process, see \cite{15}, section 8.2.5, for details.

\item[(iii)] The Diggle-Cressie-Loosmore-Ford test (see \cite{36} and \cite{spatstat}, section 10.7.4) computes a Cram\'er-von Mises type test statistic
    \begin{equation*}
    D=\int_0^R(\widehat{L}(r)-L(r))^2\mbox{d}r.
    \end{equation*}
    Here, $L(\cdot)$ is the theoretical $L$-function of a homogeneous Poisson point process, $\widehat{L}(\cdot)$ is an estimator of $L(\cdot)$,
    and $R$ is a chosen upper limit on the range of distances of interest. A Monte Carlo type test, see \cite{spatstat}, section 10.6, is then applied to $D$ to obtain a suitable rejection region.
\end{itemize}

For the simulation of alternative point processes we used the methods included in the {\tt R}-package {\tt spatstat}, as described in \cite{spatstat}.
In view of the results in Section \ref{sec:alternativ} we chose an inhomogeneous Poisson point process with intensity measure $\lambda f(x,y)\textrm{d}(x,y)$,
where $f:[0,1]^2\rightarrow [0,\infty)$ is a bounded continuous function with $\int_{[0,1]^2}f(x,y) \textrm{d}(x,y)=1$. We chose for $x,y\in[0,1]$
\begin{eqnarray*}
f_1(x,y)&=&\frac67(x+y)^2,\\
f_2(x,y)&=&\frac{2}{\sin(2)+\sin(1)-\sin(3)}\sin(2x+y),\\
f_3(x,y)&=&\frac{240}{23}((x-0.5)^2+(y-0.5)^4),\\
f_4(x,y)&=&\frac{240}{217}(1-(x-0.5)^2-(y-0.5)^4).
\end{eqnarray*}

\begin{table}[p]
\centering
\caption{Empirical rejection rates for inhomogeneous Poisson point processes, $\kappa=1$}\label{tab:inh.PPP}
\renewcommand{\tabcolsep}{1.7mm}
\small
\begin{tabular}{c|c|ccc|ccc|ccc|ccc|ccc}
\multicolumn{2}{c}{} & \multicolumn{3}{c}{$T_A$} & \multicolumn{3}{c}{$T_P$} & \multicolumn{3}{c}{$T_\chi$} & \multicolumn{3}{c}{$T_c$} & \multicolumn{3}{c}{}\\
Alt. & $\lambda\backslash c$ & 1 & 2 & 5 & 1 & 2 & 5 & 1 & 2 & 5 & 1 & 2 & 5 & $H$ & $Q$ & $D$\\ \hline
\multirow{ 4}{*}{$\mathcal{P}_{\lambda f_1}$}
& 50  & 19 & 2 & 23 & 36 & 13 & 23 & 7  & 21 & 61 & 58 & 29 & 61 & 74 & 95 & 92 \\
& 100 & 49 & 2 & 25 & 77 & 23 & 49 & 23 & 40 & 85 & 91 & 67 & 32 & 92 & $\ast$ & $\ast$ \\
& 200 & 72 & 2 & 73 & 92 & 70 & 73 & 55 & 71 & 57 & $\ast$ & 97 & 90 & 99 & $\ast$ & $\ast$ \\
& 500 & 99 & 2 & 93 & $\ast$ & 99 & 92 & 99 & 97 & 89 & $\ast$ & $\ast$ & 98 & $\ast$ & $\ast$ & $\ast$ \\
\hline
\multirow{ 4}{*}{$\mathcal{P}_{\lambda f_2}$}
& 50  & 4 & 4 & 3 & 7  & 6 & 3 & 3  & 10 & 23 & 11 & 9  & 23 & 23 & 8 & 12\\
& 100 & 7 & 3 & 1 & 18 & 4 & 8 & 4  & 11 & 39 & 20 & 12 & 2  & 28 & 16& 28\\
& 200 & 5 & 5 & 8 & 18 & 6 & 8 & 4  & 18 & 7  & 35 & 17 & 28 & 37 & 40 & 67\\
& 500 & 8 & 5 & 5 & 34 & 9 & 5 & 10 & 28 & 11 & 64 & 36 & 17 & 56& 91 & 99\\
\hline
\multirow{ 4}{*}{$\mathcal{P}_{\lambda f_3}$}
& 50  & 26     & 1 & 25 & 4      & 3      & 25 & 37     & 1  & 63 & 30     & 4      & 63 & 73 & 5 & 92 \\
& 100 & 67     & 1 & 31 & 50     & 8      & 57 & 66     & 7  & 87 & 80     & 33     & 34 & 96 & $\ast$ & $\ast$ \\
& 200 & 92     & 2 & 80 & 86     & 55     & 80 & 91     & 48 & 75 & $\ast$ & 90     & 93 & $\ast$ & $\ast$ & $\ast$ \\
& 500 & $\ast$ & 2 & 97 & $\ast$ & $\ast$ & 97 & $\ast$ & 98 & 97 & $\ast$ & $\ast$ & 99 & $\ast$ & $\ast$ & $\ast$ \\
\hline
\multirow{ 4}{*}{$\mathcal{P}_{\lambda f_4}$}
& 50  & 4 & 4 & 2 & 4 & 6 & 2 & 3  & 7 & 18 & 6 & 6 & 18 & 17 & 5  &  6\\
& 100 & 6 & 4 & 1 & 7 & 4 & 5 & 4 & 6 & 31 & 7 & 6 & 1 & 17 & 5 & 6 \\
& 200 & 5 & 5 & 5 & 6 & 5 & 5 & 4 & 6 & 4 & 8 & 6 & 19 & 16 & 6  & 10 \\
& 500 & 5 & 5 & 2 & 7 & 5 & 2 & 5 & 6 & 6 & 9 & 7 & 9 & 17 & 8 & 21 \\
\hline
\multirow{ 4}{*}{$BSP$}
& 50  & 99 & $\ast$ & 10 & 80 & $\ast$ & 10 & 0  & 91 & 41 & 97 & $\ast$ & 42 & 88 & 10  &  80\\
& 100 & $\ast$ & $\ast$ & 10 & $\ast$ & $\ast$ & 29 & 0 & 98 & 66 & $\ast$ & $\ast$ & 11 & 98 & 25 & 55 \\
& 200 & $\ast$ & $\ast$ & 37 & $\ast$ & $\ast$ & 37 & 3 & $\ast$ & 35 & $\ast$ & $\ast$ & 64 & $\ast$ & 15  & 46 \\
& 500 & $\ast$ & $\ast$ & 45 & $\ast$ & $\ast$ & 45 & 49 & $\ast$ & 59 & $\ast$ & $\ast$ & 64 & $\ast$ & 12  & 32 \\
\hline
\multirow{ 4}{*}{$MCP$}
& 50  & 49 & 21 & 37 & 55 & 26 & 37 & 25 & 15 & 66 & 70 & 42 & 66 & 87 & 59  &  71\\
& 100 & 59 & 31 & 34 & 80 & 33 & 51 & 45 & 31 & 78 & 84 & 70 & 39 & 94 & 89 & 93 \\
& 200 & 59 & 43 & 55 & 78 & 49 & 54 & 58 & 48 & 46 & 91 & 86 & 72 & 96 & 95  & 99 \\
& 500 & 64 & 52 & 52 & 81 & 60 & 51 & 76 & 61 & 55 & 95 & 96 & 67 & 96 & $\ast$  & $\ast$ \\
\hline
\end{tabular}
\end{table}

\begin{table}[p]
\centering
\caption{Empirical rejection rates for inhomogeneous Poisson point processes, $\kappa=3$}\label{tab:kappa3}
\renewcommand{\tabcolsep}{1.7mm}
\small
\begin{tabular}{c|c|ccccc|ccccc|ccccc|ccccc}
\multicolumn{2}{c}{} & \multicolumn{5}{c}{$T_A$} & \multicolumn{5}{c}{$T_P$} & \multicolumn{5}{c}{$T_\chi$} & \multicolumn{5}{c}{$T_c$} \\
Alt. & $\lambda\backslash c$ & 1 & 2 & 3 & 4 & 5 & 1 & 2 & 3 & 4 & 5 & 1 & 2 & 3 & 4 & 5 &1 & 2 & 3 & 4 & 5\\ \hline
\multirow{ 4}{*}{$\mathcal{P}_{\lambda f_1}$}
& 50   & 57 & 51 &  4 & 1 & 4 & 7 & 12 & 37 & 4 & 0 & 16 & 6 & 1 & 0 & 1 & 77 & 60 & 18 & 12 & 25 \\
& 100  & 94 & 82 & 46 & 10 &  0 & 17 &  3 & 51 & 71 & 31 & 25 &  7 & 10 & 0 &  0 & 96     & 96    & 88 & 57 & 37\\
& 200  & 99 & 97 & 36 &  1 & 16 & 25 & 23 & 98 & 91 &  3 &  9 & 30 &  8 & 8 & 65 & $\ast$ & $\ast$& 99 & 94 & 92\\
& 500  & $\ast$ & $\ast$ & 99 & 16 & 9  & 77 & 12 & $\ast$ & $\ast$ & 95 & 9 & 77 & 81  & 0 & 95 & $\ast$ & $\ast$ & $\ast$ & $\ast$ & $\ast$ \\
\hline
\multirow{ 4}{*}{$\mathcal{P}_{\lambda f_2}$}
& 50   & 7  & 9  & 3 & 6 & 6 & 3 & 2 & 6 & 3 & 2 & 9 & 2 & 3 & 1 & 3 & 15 & 9  & 5 & 7  & 9 \\
& 100  & 20 & 6 & 4 & 4 & 3 & 3 & 2 & 16  & 12 & 5 & 9 & 4 & 4 & 1 & 2 & 38 & 25 & 18 & 10  & 10 \\
& 200  & 11 & 10 & 3 & 3 & 6 & 3 & 2 & 16 & 12 & 5 & 2 & 5 & 3 & 13 & 13 & 45 & 49& 34 & 22 & 19 \\
& 500  & 33 & 20 & 9 & 4 & 6 & 7 & 8 & 52 & 52 & 6 & 1 & 23 & 7  & 7  &43  & 70 & 85 & 80 & 63 & 42 \\
\hline
\multirow{ 4}{*}{$\mathcal{P}_{\lambda f_3}$}
& 50   & 62 & 64 & 3 &0 &11  & 81 & 9 & 1 & 0 & 1 & 64 & 58 & 39 & 15 & 18 & 66 & 63 & 7 & 0 & 15\\
& 100  & 98 & 91 & 60 & 8 & 0 & 98 & 67 & 0 & 2 & 0 & 70 & 46 & 87 & 6 & 2 & 92 & 94 & 65 & 3 & 0\\
& 200  & $\ast$& $\ast$ & 46& 0 & 29 & 99& 0& 85& 51& 0& 11& 80&37 &0 &8 &$\ast$ &$\ast$ &89 &44 &70 \\
& 500  & $\ast$ & $\ast$& $\ast$& 14& 19& $\ast$& 1& $\ast$ & $\ast$&  71& 8 & 98& 99& 0& 92 &$\ast$ & $\ast$&$\ast$ &$\ast$ &$\ast$ \\
\hline
\multirow{ 4}{*}{$\mathcal{P}_{\lambda f_4}$}
& 50   & 3& 6& 3& 6& 5& 4& 2& 4& 4& 4& 12& 3& 5& 2& 4& 7& 6& 4& 6& 7\\
& 100  & 8& 3& 3& 5& 4& 3& 3& 8& 5& 6& 11& 7& 6& 3& 3& 14& 7& 6& 6& 7\\
& 200  & 3& 5& 4& 4& 4& 3& 4& 7& 7& 4& 2& 4& 4& 6& 5& 9& 10& 9& 8& 9\\
& 500  & 5& 4& 5& 5& 5& 5& 6& 9& 10& 6& 1& 8& 4& 4& 10& 11&13 &12 &11 &10 \\
\hline
\end{tabular}
\end{table}

We also considered a further alternative point process, namely the Baddeley-Silverman cell process $BSP$, as proposed in \cite{32}. This process is designed to have the same second-order properties as an homogeneous PPP so that it cannot be detected by $K$- or $L$-function methods. It's the standard counterexample to the claim that these functions completely characterize the point pattern. The $BSP$ is generated by dividing the observation window into equally spaced quadrats in which either $0, 1$ or $10$ points are independently uniformly scattered with probabilities $1\slash10, 8\slash9$ and $1\slash90$ respectively.

Due to ongoing interest in the detection of clusters in point patterns we considered the special case of a Neyman-Scott cluster process (for details see \cite{15}, p. 662), namely the Mat\'ern cluster process $MCP$, see \cite{spatstat}, p. 139. In this model one first simulates a homogeneous PPP as parent points with fixed intensity $\widetilde{\lambda}$. In a second step, one generates  a Poisson random number $\widetilde{N}$ of independently uniformly distributed points in a disc of radius $r$ center around each parent point. Discarding the parent process yields the $MCP$. We chose $\widetilde{\lambda}:=\lambda\slash m, r=0.2$ and set the parameter of the Poisson distribution of $\widetilde{N}$ to $m=\scriptstyle\left\lfloor\sqrt{\lambda\slash\kappa}\right\rfloor$.

Table \ref{tab:inh.PPP} and Table \ref{tab:kappa3} show the percentages (out of 10~000 replications) of rejections of $H_0$ rounded to the nearest integer. In both tables, it is obvious that $T_j$, $j\in\{A,P,\chi\}$, as well as $T_c$ depend crucially on the choice of the threshold parameter $c$. The difference between Table \ref{tab:inh.PPP} and Table \ref{tab:kappa3} is the choice of $\kappa$, which is controlled by the number $m^2$ of bins considered, and the simulation results clearly show the impact of a proper choice of $c$ and $m$. Fixing $\kappa=1$ as in Table \ref{tab:inh.PPP} a natural choice of the threshold parameter is $c=1$, for which $T_1$ gives the best overall performance of the new methods and competes with the compared tests. In case of the BSP it outperforms the $\chi^2$-quadrat-count-test as well as the Diggle-Cressie-Loosmore-Ford-test. Table \ref{tab:kappa3} provides more insight into the power of the new tests and the dependence on the parameters. Since $\kappa$ is the mean number of points falling into a single bin, one would expect the best performance for $\kappa=c$. Throughout the table, this assumption does not seem to hold. A slightly lower threshold parameter than $\kappa$ results in higher power of the tests. The power of the testing procedure that combines more than one Minkowski functional outperforms overall the procedures based on a single functional. Notice that the presented simulation results are based on a naive choice of the parameters, so there is hope to find better performing tests by optimizing the choice of $\kappa$ and $c$, preferably in a data driven way.


\section{Data analysis of gamma-ray astronomy}\label{sec:gamma-ray}

As an exemplary application, we analyze experimental observations from astroparticle physics, the gamma-ray sky map of the {\em Fermi Gamma-ray Space Telescope}.
This space observatory explores extreme astrophysical phenomena, namely high-energetic radiation from both galactic and extragalactic sources.
These sources may be fast rotating neutron stars or hot gas moving nearly at the speed of light.
The satellite was launched in 2008 and detects the gamma-rays from a low Earth orbit.
The energies of the singly detected photons range from 20\,MeV to 300\,GeV; for comparison, a photon with a visible wavelength has energies below 3\,eV.

Here we analyze a sky-map of gamma-rays observed by the so-called {\em Large Area Telescope} (LAT)~\cite{0004-637X-697-2-1071}.
The directions of the incoming photons form a point pattern on the hemisphere.
If the observation window is restricted to a small field of view, the curved space can be well approximated by a rectangular observation window in the Euclidean plane.
If necessary, the analysis could easily be adapted from the plane to the sphere.

\begin{figure}[t]
  \centering
  \includegraphics{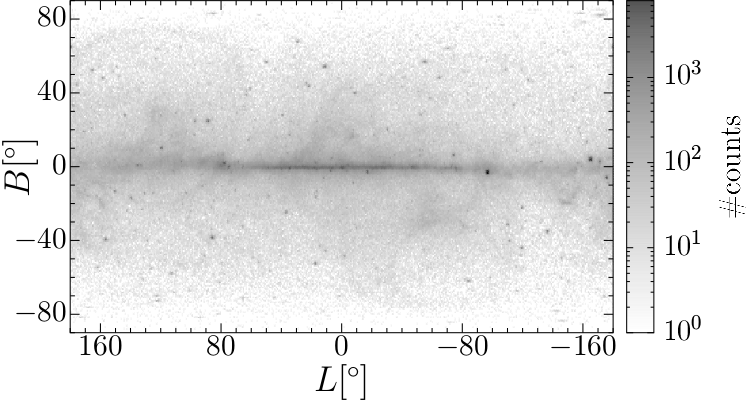}
  \caption{A gamma-ray sky map as recorded by the {\em Large Area Telescope}.
     The binned map of the whole sky is based on single gamma-like events with energies above 2\,GeV.
  Point-like sources of gamma-rays and a diffuse emission from our galactic plane are clearly visible.
  The events can be modeled by a inhomogeneous Poisson point process.}
  \label{fig:Fermi-sky-map}
\end{figure}

Figure~\ref{fig:Fermi-sky-map} shows a binned sky map, that is, the gray values represent the numbers of events detected within each bin.
We here analyze a data set of $10^6$ events (which were collected within about two months).
The positions in the sky are given in galactic coordinates, that is, the galactic latitude $B$ and the galactic longitude $L$.
The supermassive black hole in the center of our galaxy is at $L=0$ and $B=0$.
Above and below the black hole, there is an unresolved phenomenon, the so-called {\em Fermi-Bubbles}, as can be seen in Fig.~\ref{fig:Fermi-sky-map}.
Its origin is still unknown, but it might be the result of a more active period of the supermassive black hole in the center of our galaxy, see \cite{BordoloiEtAl2017}.

The null hypothesis is well-defined for any choice of the bin width, because the Poisson point process does not assume an intrinsic length.
However, for the application to real data, a reasonable choice of the bin width in accordance with the resolution of the detector improves the sensitivity of the analysis.
A too coarse graining can hide distinct geometrical features of sources within a single pixel. If the mesh is too fine compared to the point spread function, the black and white image can easily be dominated by the random scattering of the signals.

The resolution of the LAT depends strongly on the energies of the photons. It resolves the direction of the photons within a few degrees for 1\,MeV gamma-rays  but with a resolution of about 0.2 degrees for energies above 2 GeV, see \cite{AceroEtAl2015}.
The energy of the events analyzed here ranges from 1\,GeV to 1\,TeV. Accordingly, we choose a bin width of about 0.2 degrees or larger.

\begin{figure}[t]
 \centering
 \includegraphics[scale=0.5]{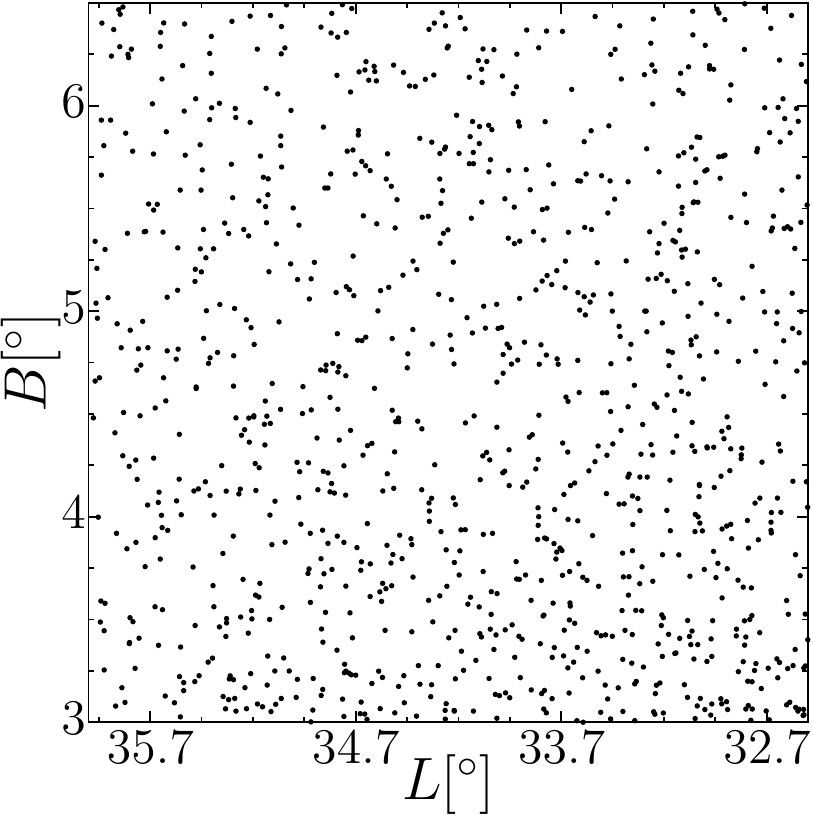}\hspace{0.5cm}\includegraphics[scale=0.5]{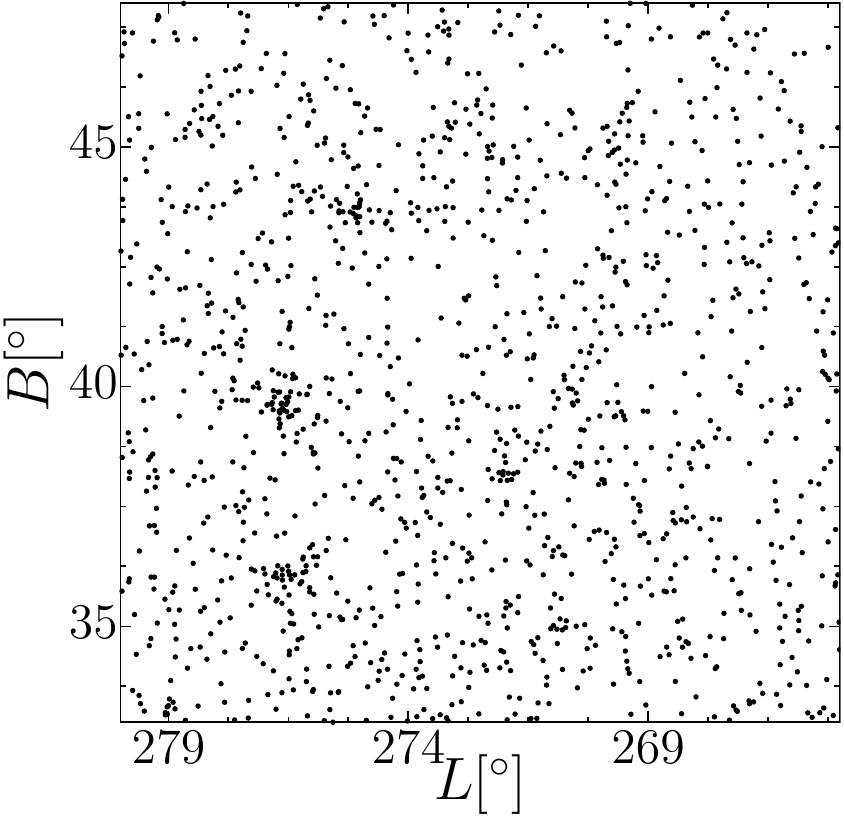}\hspace{0.5cm}\includegraphics[scale=0.5]{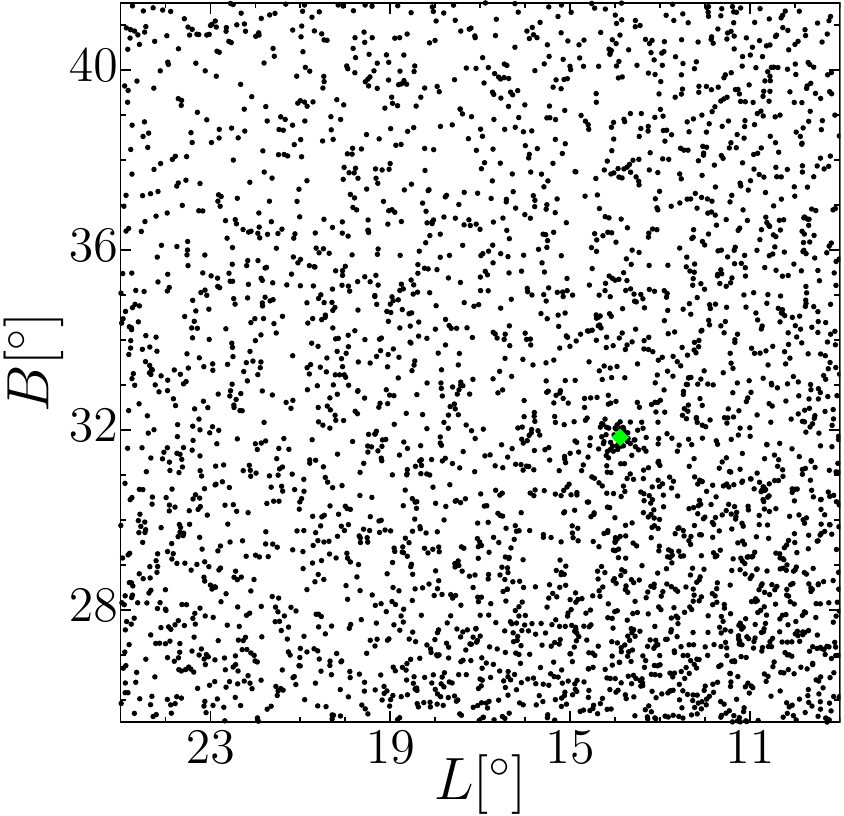}\\[2mm]
 \includegraphics[scale=0.67]{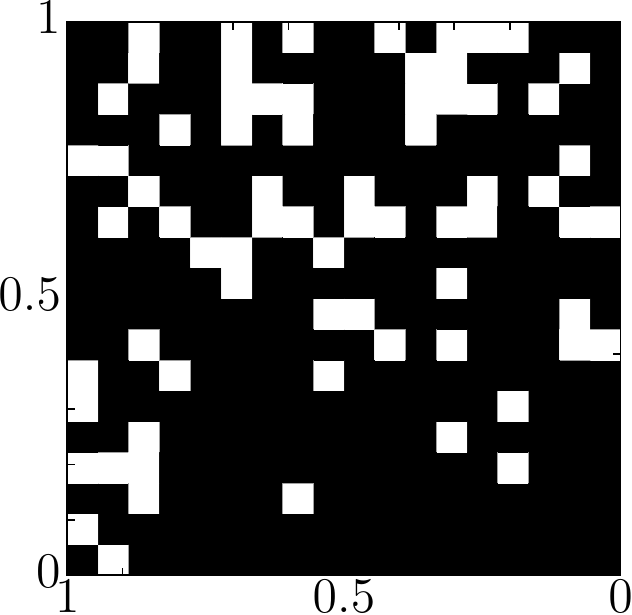}\hspace{0.5cm}\includegraphics[scale=0.67]{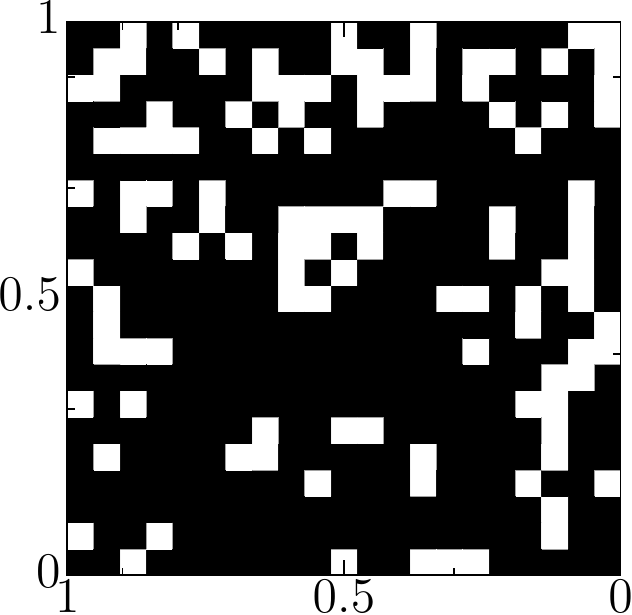}\hspace{0.5cm}\includegraphics[scale=0.67]{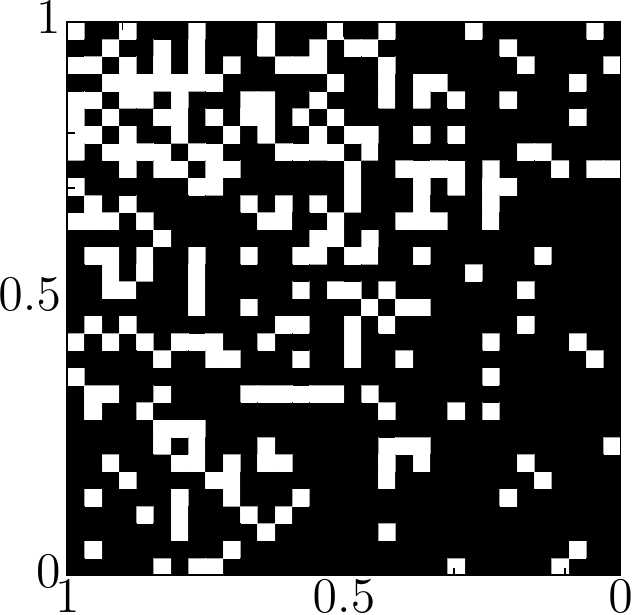}\\[2mm]
  \hspace{0.5cm}(1)\hspace{4cm}(2)\hspace{4.5cm}(3)
  \caption{Excerpts from the observed gamma-ray sky map (data sets 1-3): the point patterns formed by the single events (top) and the resulting binary images for $c=2$ (bottom).}
  \label{fig:Band-Point}
\end{figure}

There are four contributions of the events detected by the LAT:
\begin{enumerate}
  \item strong galactic point sources, like active galaxies of the blazar class or pulsars~\cite{AceroEtAl2015},
  \item diffuse radiation from galactic gas clouds, that is produced by collisions of high energetic protons and particles in the gas clouds,
  \item an isotropic diffuse background from extragalactic sources, that cannot be resolved as single sources, but adds up to a homogeneous background, and
  \item an isotropic background of high-energetic protons. A small fraction of the protons in the cosmic radiation are incorrectly but unavoidably classified as gamma-rays.
    They arrive homogeneously from all directions after a diffusion in the galactic magnetic fields.
\end{enumerate}
From a statistical point of view, these four contributions can be interpreted as:
\begin{enumerate}
  \item a strongly clustering (Mat\'ern-type) point process,
  \item an inhomogeneous Poisson point process,
  \item and d) a homogeneous Poisson point process.
\end{enumerate}
Here we analyze field of views that contain point sources within a homogeneous background as well as inhomogeneous diffusive radiation from the galactic disc,
see Fig.~\ref{fig:Band-Point}.
%
The first data set (upper left picture in Figure \ref{fig:Band-Point}) contains 1041 point events.
We chose as parameters $\lambda=1041$ and $m=18$, which results in $\kappa\approx3.213$.
The second data set (upper middle picture in Figure \ref{fig:Band-Point}) contains 1339 point events ($\lambda=1339$, $m=21$ and $\kappa\approx3.036$) and the third data set (upper right picture in Figure \ref{fig:Band-Point}) exhibits 3193 point events ($\lambda=3193$, $m=32$ and $\kappa\approx3.118$). For all three analysis the thresholding parameter $c$ was fixed to 2 in accordance with the insight gained in section \ref{sec:simul}. The third data set is a region in the northern Fermi bubble with diffuse radiation from gas clouds, which causes a global gradient in the point pattern, and a point source, which is marked by the diamond symbol ($\diamond$). The latter is listed in the LAT 4-year Catalog as the source J1625.1---0021~\cite{AceroEtAl2015}, which is likely to be a millisecond pulsar, e.g., see~\cite{DaiEtAl2016}. For computing the $p$-values in Table \ref{tab:real.data} we used the asymptotic $\chi^2$ distribution with 1 and 3, degrees of freedom, respectively.
\begin{table}[t]
\centering
\caption{Calculated $p$-values for the data sets of Figure \ref{fig:Band-Point}}\label{tab:real.data}
\renewcommand{\tabcolsep}{1.7mm}
\begin{tabular}{c|cccc|c}
Dataset & $T_A$ & $T_P$ & $T_\chi$ & $T_c$ & $H$ \\ \hline
1 & 2.560e-2 & 2.134e-1 & 4.260e-1 & 2.550e-2 & 2.616e-1 \\
2 & 1.389e-3 & 2-927e-1 & 1.501e-2 & 2.900e-5 & 1.069e-1 \\
3 & 2.225e-6 & 1.369e-3 & 4.507e-1 & 9.312e-8 & 2.474e-5 \\ \hline
\end{tabular}
\end{table}
Table \ref{tab:real.data} shows that $T_c$ rejects the hypothesis of CSR for all 3 data sets, while $T_P$, $T_\chi$ and $H$ clearly fail to detect the alternative in data set 1 even for a larger level $\alpha$ like $0.1$. Data set 2 indicates the gain of power by considering the combination of 3 functionals, compared to the single functional. For data set 3 all tests, except $T_\chi$, detect the inhomogeneous radiation in the region of the Fermi bubble. For reference we included the $p$-values of the Hopkins and Skellam test. The quadrat count and Diggle-Cressie-Loosmore-Ford tests reject the null hypothesis, but since their power depend on parameters and on the number of Monte Carlo replications the $p$-values are omitted since the comparability is questionable.

To detect point sources, the diffuse background radiation is estimated based on maps of galactic gas clouds.
These estimates are then subtracted from the data.
However, because of limited observation of the gas clouds and the complex interactions between the high-energetic protons with the gas, systematic effects remain that may hide point sources in regions of strong diffuse emission. 

A new approach to distinguish the signals of such hidden point sources could detect currently undetected sources in the same data.
Varying the threshold, we can in principle separate the detection of point sources and diffuse emission.
Moreover, our test needs no a-priori assumptions about the complex shape of the gas clouds, but nevertheless the test statistic includes geometric information of the sources in the field of view.

Here we have only applied our method as a proof of principle for a rigorous morphometric null-hypothesis test in gamma-ray astronomy.
With further optimizations for applications in astroparticle physics, our approach could have the potential to detect new gamma-ray sources and help to unravel some unknown phenomena.

\section{Further Comments and Conclusions}\label{sec:comm.conc}
Our new tests are presented for 2-dimensional data sets due to the look-up Table \ref{tab:MinkowskiFunctionalsLookUpTable} for Minkowski functionals. Nevertheless the table can straightforwardly be generalized to Minkowski functionals in higher dimensions in the following way. In $d$-dimensional Euclidean space ($d>2$), the $(2\times2)$-neighborhood must be extended to a $2^d$-neighborhood. The total number of local configurations is therefore $2^{2^d}$. For each configuration, the corresponding entry in the look-up table of a Minkowski functional is given by an explicit limit of integrals. Starting with the Minkowski sum of the black pixels with a ball of radius $\varepsilon$, that is, the parallel body of the interior, an intersection with the interior of the $2^d$-neighborhood yields a smooth body for which the integral representation of the Minkowski functionals can be calculated, see \cite{43}. The limit $\varepsilon\rightarrow 0$ yields the entry in the look-up table. For the volume, this is equal to the number of black pixels divided by $2^d$, and the contribution to the surface area is given by the number of neighboring pairs of black and white pixels, divided by $2^{d-1}$. The mean width is determined by the opening angles of the edges of black pixels that are neighbors of white pixels. Similarly the local contributions to the Euler characteristic can be expressed by the corners between black and white pixels.

Furthermore, our analysis can also be applied to real data that are distorted by detector effects, like a varying camera acceptance. Instead of an initially homogeneous Poisson point process, the recorded data then follows an inhomogeneous Poisson point process with a known intensity function. Such detector effects can easily be corrected by adding Monte Carlo Poisson events or by performing a Monte Carlo post-selection~\cite{Goering2012,24}. Under the null-hypothesis for the initial data, the resulting post-processed data is again a stationary Poisson point process. If the corrections are applied locally, we can compensate even a strong suppression of signals or subtract strong known point sources~\cite{44}.

We want to indicate some open problems related to the tests. Throughout the article, we assume that the intensity $\lambda$ of ${\cal P}_\lambda$ is known, so an interesting question is what effect an estimator $\widehat{\lambda}$ has in the theoretical derivations of the tests. Section \ref{sec:alternativ} describes the behaviour of the Minkowski functionals under fixed alternatives, but it is still unknown the results will lead to statements regarding consistency of the tests aigainst the inhomogeneous Poisson point process and is totally open for point process alternatives with inherent dependency structure. The simulation study suggests that finding a best (data dependent) choice of the parameters $c$ and $m$ is crucial to increase the power of the tests. The presented tests contain very nice features, like very fast computation time (even for big data) and flexibility with regard to the choice of parameters, which can lead to better power performance for special alternatives.

Finally, we emphasize that the approach to analyze point patterns by means of Minkowski functionals of binary images might be the starting point of an rich and interesting path to follow for further research, and therefore give some examples.
\begin{itemize}
\item Simulations show that one may obtain better performing procedures by looking at more than one threshold parameter $c$. We suggest to investigate $\max_{c\in\N} T_c$ or $\sum_{c=1}^\infty T_c$ for which we expect procedures with distinctly higher power and a greater flexibility in detecting point-like or extended sources, for first simulations see \cite{44}.

\item To detect a local deviation of CSR, we suggest to use the presented tests in a moving window approach, see \cite{42}.

\item In view of Hadwiger's characterization theorem one might find a linear combination of the Minkowski functionals that is most powerful against alternatives which are additive, continuous and invariant under rigid motions.

\item Tensorial Minkowski functionals are generalizations of Minkowski functionals (also called scalar Minkowski functionals), see \cite{19,43}. They directly quantify the degree of anisotropy and the preferred orientation in an anisotropic system. One can easily derive a corresponding look-up table \cite{44} and define analogous tests to better detect anisotropic deviations from a Poisson point process.

\item The formulae given in Theorem \ref{varMF} and \ref{covarMF} can also be used to analyze the non-occurrence of dependency in binary image (or boolean matrix) data directly by means of Minkowski functionals.
\end{itemize}

\section*{Acknowledgements}
The authors want to thank M. Penrose for indicating the reference \cite{20}. We appreciate the data provided by the Fermi Gamma-Ray Spactelescope and thank Stefan Funk, director of the Erlangen Center for Astroparticle Physics (ECAP), for his insights into gamma-ray astronomy and the Fermi telescope, his assistance with the data of the Fermi space telescope, and very helpful discussions and advice.

\bibliographystyle{apalike}

\bibliography{literatur}
\end{document}